\documentclass[twocolumn, notitlepage, 10pt, aps, floatfix, showpacs, prb, citeautoscript, superscriptaddress]{revtex4-1}
\usepackage{graphicx}
\usepackage{amsmath, amssymb}
\usepackage{bm}
\usepackage{xcolor}
\usepackage[colorlinks, citecolor=blue, urlcolor=blue, linkcolor=red]{hyperref}
\usepackage{bookmark}
\usepackage[version=4]{mhchem}
\usepackage{microtype}
\usepackage{float}
\usepackage[load=physical,load=abbr, range-units=single]{siunitx}

\begin{document}
\title{Presence versus absence of end-to-end nonlocal conductance correlations in Majorana nanowires: Majorana bound states versus Andreev bound states}

\author{Yi-Hua Lai}
\author{Jay D. Sau}
\author{Sankar Das Sarma}
\affiliation{Department of Physics, Condensed Matter Theory Center and the Joint Quantum Institute, University of Maryland, College Park, MD 20742}

\begin{abstract}
By calculating the differential tunneling conductance spectra from the two ends of a Majorana nanowire with a quantum dot embedded at one end, we establish that a careful examination of the nonlocal correlations of the zero bias conductance peaks, as measured separately from the two ends of the wire, can distinguish between topological Majorana bound states and trivial Andreev bound states. In particular, there will be identical correlated zero bias peaks from both ends for Majorana bound states, and thus the presence of correlated zero bias conductance from the two wire ends could imply the presence of topological Majorana zero modes in the system. On the contrary, there will not be identical correlated zero bias peaks from both ends for Andreev bound states, so the absence of correlated zero bias conductance from the two wire ends implies the absence of topological Majorana zero modes in the system. We present detailed results for the calculated conductance, energy spectra, and wavefunctions for different chemical potentials at the same magnetic field values to motivate end-to-end conductance correlation measurements in Majorana nanowires.
\end{abstract}

\maketitle

\section{Introduction}\label{sec:level1_1}
It was pointed out by Kitaev \cite{Kitaev2001Unpaired} that isolated Majorana zero modes (i.e. topological bound states with precise zero energy) existing at the ends of a one dimensional (1D) spinless p-wave superconductor are effective non-Abelian anyons which could potentially be used for fault-tolerant topological quantum computation \cite{Nayak2008NonAbelian,Sarma2015Majorana,Lutchyn2018Majorana}. It was also theoretically established around the time of Kitaev's work that such Majorana zero modes in a topological p-wave superconductor can be detected by the usual normal metal-superconductor (NS) differential tunneling conductance measurements which would reflect a quantized zero bias conductance peak (ZBP or ZBCP) associated with the Majorana bound states (MBS) \cite{Sengupta2001Midgap}. The fact that the presence or absence of a quantized ZBCP in the NS tunneling spectroscopy signals the presence or absence of MBS was later rediscovered and expanded on by several groups \cite{Law2009Majorana,Flensberg2010Tunneling,Akhmerov2009Electrically,Lin2012Zerobias,Ben-Shach2015Detecting}. The subject took on particular significance after it was shown theoretically that two-dimensional and 1D semiconductor structures could actually host MBS under well-defined experimentally achievable conditions \cite{Sau2010Generic,Sau2010NonAbelian,Tewari2010theorem,Lutchyn2010Majorana,Oreg2010Helical}, and soon after these concrete predictions for the possible existence of MBS in low-dimensional semiconductor structures, Mourik et al. reported \cite{Mourik2012Signatures} the experimental observation of a ZBP in the tunneling conductance of 1D InSb nanowires (on NbTiN superconducting substrates) loosely consistent with the theoretical predictions. This started a deluge of theoretical and experimental activity, which continues unabated for the last 7 years, in semiconductor (InSb and InAs) nanowires (with NbTiN and Al as the parent superconductor) aimed at the observation and elucidation of ZBCPs which are considered to be the signatures for the putative MBS in these Majorana nanowires. The subject got particular impetus from the important backing of Microsoft Corporation, which started a large technological development effort in building a topological quantum computer based on these semiconductor Majorana nanowires \cite{Castelvecchi2017Quantum}.

In spite of enormous experimental progress \cite{Das2012Zerobias,Deng2012Anomalous,Churchill2013Superconductornanowire,Finck2013Anomalous,Deng2016Majorana,Chen2017Experimental,Nichele2017Scaling,Zhang2017Ballistic,Gul2018Ballistic,Zhang2018Quantized} in materials fabrication leading to the ubiquitous observation of impressive ZBCPs in Majorana nanowires in many laboratories far surpassing the quality of the ZBCPs originally reported by Mourik et al. \cite{Mourik2012Signatures}, questions, however, linger on whether MBS (as opposed to mere ZBPs in the tunneling measurements) have actually been seen yet. In particular, Ref.~\cite{Liu2017Andreev} forcefully raised the key question on whether many, if not all, of the experimentally observed ZBCPs in Majorana nanowires could have originated from accidental non-topological (often called 'trivial' in this context) Andreev bound states (ABS) which fortuitously happen to reside near zero energy inside the superconducting gap. These trivial almost-zero-energy midgap ABS could be producing the ZBPs in the experiments fooling everybody into thinking that MBS have been observed whereas in reality what have been observed are the ZBP associated with these non-topological ABS. Actually, the possibility that there could be generic low-lying in-gap fermionic bound states in Majorana nanowires arising from impurity disorder \cite{Liu2012ZeroBias,Bagrets2012Class,Pikulin2012zerovoltage,Mi2014Xshaped,Sau2013Density} and/or inhomogeneous  chemical potential \cite{Kells2012Nearzeroenergy,Prada2012Transport,Liu2017Andreev,Moore2018Twoterminal,Moore2018Quantized,Vuik2018Reproducing} was pointed out early in the literature, and the fact that well-defined ZBCP could arise from low-lying ABS in confined semiconductor structures was also experimentally demonstrated \cite{Lee2012ZeroBias}. If topological MBS and trivial ABS could both produce tunneling ZBCP in Majorana nanowires, a serious problem arises in the interpretation of the experimental data. One cannot automatically assume the experimental observation of a ZBCP as evidence for the existence of an underlying isolated MBS. This is particularly true in light of the fact that low-energy midgap ABS seem to be generic in Majorana nanowires arising from chemical potential inhomogeneity or isolated impurities acting as quantum dots in the system. The interplay of spin-orbit coupling and Zeeman splitting generically allows Andreev bound states to reside close to zero energy over finite ranges of the external magnetic field. The ABS thus mimic the zero-energy MBS with the big difference that MBS arise as zero energy modes in the nanowire only in the topological regime (which translates to the induced Zeeman spin splitting being larger than the critical field necessary for the topological quantum phase transition, TQPT), whereas the ABS arise in the trivial regime for Zeeman field below the critical field. Experimentally, unfortunately, there is no independent way of precisely knowing the critical field, so one does not apriori know whether an observed ZBCP happens to be in the topological or trivial regime. Thus, although the existence of a ZBCP may be a necessary condition for the existence of underlying MBS, it is by no means sufficient since the almost zero-energy ABS produces similar ZBCP in the nontopological regime. The inability to decisively distinguish between ZBCPs arising from MBS and ABS has been the crucial stumbling block in further progress in the subject. 

Given the great importance of the distinction between ABS and MBS to the subject of Majorana nanowires, it is understandable that many theoretical papers have appeared following Ref.~\cite{Liu2017Andreev} with various proposals on how to discern MBS from ABS \cite{Chiu2017Conductance,Setiawan2017Electron,Moore2018Twoterminal,Liu2018Distinguishing,Moore2018Quantized,Stanescu2018Illustrated}. The situation has, however, remained unclear, and even the unambiguous observation \cite{Zhang2017Ballistic} of the predicted quantized Majorana ZBCP can be interpreted in terms of underlying ABS accidentally localized at midgap \cite{Stanescu2018Illustrated}. This is the context of the current work where we propose a definitive experiment which, in principle, can distinguish between ABS and MBS based on the extensively used conductance tunneling spectroscopy. Our idea is surprisingly simple and basic, and it is therefore somewhat puzzling that this idea has not been discussed in details in the vast existing nanowire literature in the context of distinguishing between MBS and ABS.

We show that a careful comparison between the conductance spectra obtained by carrying out  tunneling measurements from the two ends of a Majorana nanowire should be able to decisively distinguish between ZBCPs arising from MBS and ABS. In particular, ZBCPs from MBS (ABS) would manifest highly correlated (uncorrelated) low-energy behavior. We provide extensive numerical simulations to demonstrate the importance of simultaneous tunneling measurements from both ends in the context of MBS versus ABS distinction. The specific system we consider is motivated by the experiment of Deng et al. \cite{Deng2016Majorana}, who carried out tunnel conductance measurements in a Majorana nanowire with an embedded quantum dot at the wire end where the NS tunnel barrier resides. They found well-defined conductance features away from zero bias at low magnetic field values, which merged at zero bias at higher magnetic field values producing sharp ZBCP. This was interpreted by the authors as evidence in favor of  trivial ABS (at finite energy) transforming into zero-energy topological MBS as the magnetic field sweeps through the TQPT. The experiment of Deng et al. was critically reanalyzed by Liu et al \cite{Liu2017Andreev} who showed that most likely the experiment is demonstrating the existence of low-energy midgap ABS, induced by the quantum dot, which is producing the ZBCP rather than the transformation of trivial ABS into topological MBS as envisioned in Ref.~\cite{Deng2016Majorana} Of course, the possibility that some of the observed ZBCP in the experiment arise from MBS cannot be ruled out theoretically, but the real problem is that there is no way to know apriori which ZBCP arise from ABS and which ones from MBS. 

Our current work establishes that an experiment similar to that in Ref.~\cite{Deng2016Majorana} with the tunneling spectroscopy carried out from both ends of the wire in the same sample (with a quantum dot only at one end) can distinguish between the ZBCPs arising from MBS and ABS through a simple examination of the correlations (or not) between two sets of tunneling data. The ZBCP arising from MBS (ABS) will be (un)correlated between the two ends--if the same ZBCP shows up in the tunneling from both ends it is likely to be associated with MBS, whereas if the ZBCP exists only in the tunneling from one end (but not the other), then it is likely to be arising from ABS. Essentially, all one needs is redoing the Deng experiment by having NS tunneling from both ends of the wire. For completeness,  we also briefly consider the effect of embedded end quantum dots on the cross-conductance, which can also be measured in the same set-up.   Such measurements have recently been proposed as a way to distinguish ABSs versus MZMs by detecting the TQPT \cite{Rosdahl2018Andreev}.

We note, in order to avoid any confusion in the terminology, that our work is on conductance correlations and not on current correlations. In particular, we are not discussing 'noise correlation' measurements in the current work-- we are discussing correlations between measured conductances from the two ends of the nanowire.

The rest of this paper is organized as follows. In Sec.~\ref{sec:level1_2}, we describe our model, theory, and calculations. In Sec.~\ref{sec:level1_3}, we present our numerical results for the calculated tunnel conductance from both ends of the wire and provide discussions on how such correlated tunneling spectroscopy could resolve the ABS versus MBS conundrum. We conclude in Sec.~\ref{sec:level1_4} providing a summary and commenting on the experimental prospects. An Appendix provides detailed numerical conductance results for different chemical potentials and magnetic fields along with the corresponding MBS or ABS wavefunctions and energy spectra.

\section{Model, Theory, and Calculations}\label{sec:level1_2}
\begin{figure}
	\includegraphics[scale=0.14]{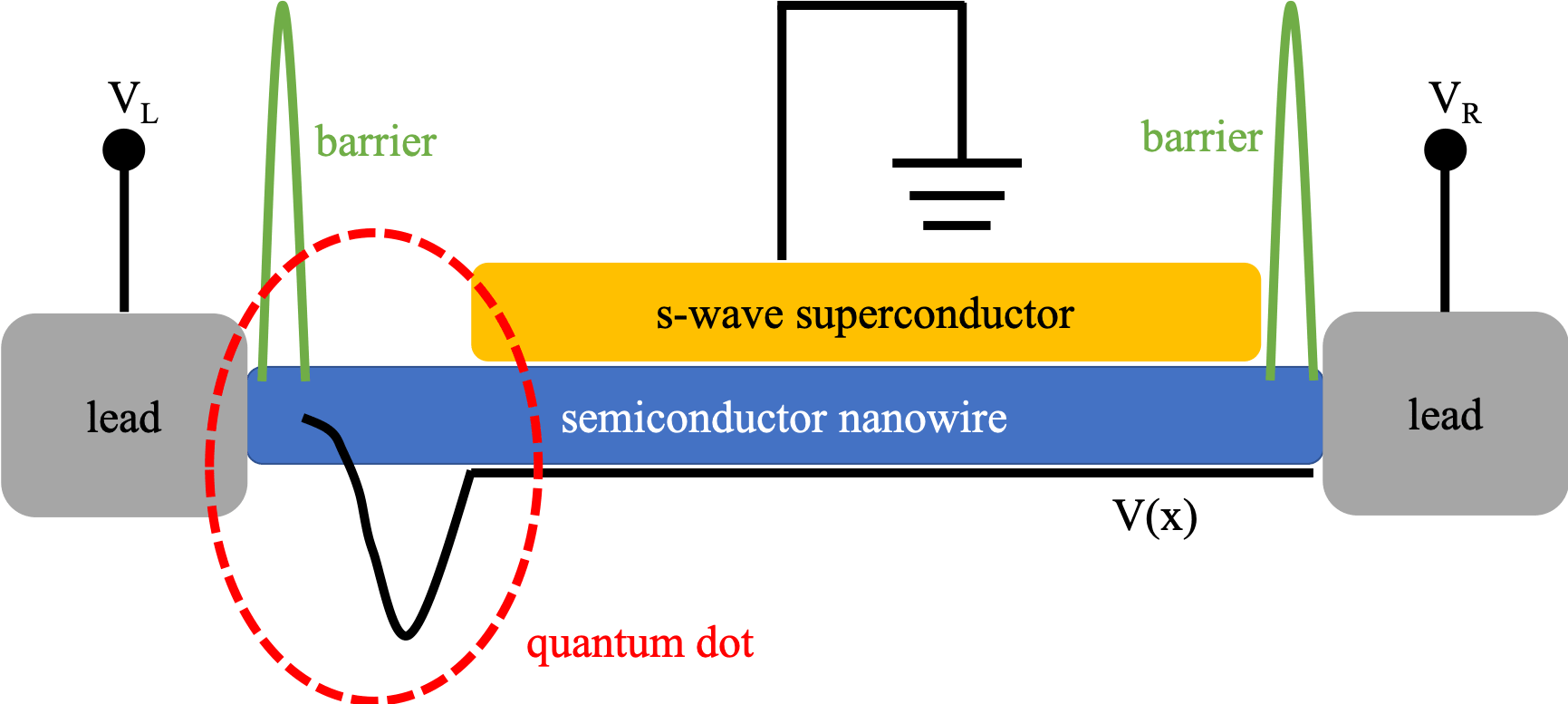}
	\caption{A schematic plot of the junction composed of leads on both sides of the quantum dot-nanowire-superconductor hybrid structure, which is basically motivated by Deng \textit{et al.} experimental setup \cite{Deng2016Majorana}. The semiconductor nanowire is covered by a parent $s$-wave superconductor, except for the part embedded by the quantum dot (shown in the red dashed line). Note that the quantum dot is strongly coupled to the nanowire, which may not exhibit Coulomb blockade effect. The leads used to measure the conductance are set on both sides, which also induce the tunnel barriers (green parts). The conductances calculated in this paper are probed either from the left lead or from the right lead.}
	\label{fig:scheme}
\end{figure}

We analyze the conductance $G_\alpha=dI_\alpha/dV_\alpha$, where $\alpha=L,R$ corresponding to left or right lead of the Majorana nanowire device (i.e. a semiconductor nanowire in the presence of proximity induced superconductivity, intrinsic Rashba spin-orbit coupling, and external magnetic field induced Zeeman spin splitting) in a set-up shown in Fig.\ref{fig:scheme} (a quantum dot is embedded at the left end) and $I_\alpha$ and $V_\alpha$ denote the current and voltage in lead $\alpha$. We model the device within a minimal single-band model \cite{Sau2010Generic,Lutchyn2010Majorana,Oreg2010Helical} described schematically by the Bogoliubov-de Gennes (BdG) Hamiltonian 
\begin{equation}\label{E1}
	\begin{aligned}
	&\hat{H}=\frac{1}{2}\int dx\hat{\Psi}^\dagger(x) H_{NW}\hat{\Psi}(x),\\
	&H_{NW}=\left(-\frac{\hbar^2}{2m^*}\partial_x^2-i\alpha_R\partial_x\sigma_y-\mu\right)\tau_z\\
	&\qquad\qquad+V_z\sigma_x+\Delta(V_z)\tau_x-i\Gamma,
	\end{aligned}
\end{equation}
where $\hat{\Psi}=(\hat{\psi}_\uparrow,\hat{\psi}_\downarrow,\hat{\psi}_\downarrow^\dagger,-\hat{\psi}_\uparrow^\dagger)^T$ is the wave function in Nambu space, $\sigma_{x,y,z}$ ($\tau_{x,y,z}$) are Pauli matrices in spin (particle-hole) space. The electrons are assumed to have an effective mass, $m^*$, Rashba spin-orbit coupling $\alpha_R$, and a magnetic field-induced Zeeman splitting $V_z$. The superconducting pairing potential, $\Delta(V_z)$, is assumed to be suppressed by Zeeman splitting $V_z$ as: 
\begin{equation}\label{E2}
	\Delta(V_z)=\Delta_0\sqrt{1-\left(V_z/V_c\right)^2},
\end{equation}
where $\Delta_0$ is the original induced superconducting gap without the magnetic field background, and $V_c$ is the field where the superconducting gap collapses. One may think of this collapse of the bulk gap as arising from the Clogston effect due to the bulk spin polarization in the parent superconductor (and other effects). Experimentally, such a bulk gap collapse is always present, and $V_c$ is a phenomenological parameter defining this field for the bulk gap collapse in the theory. A phenomenological dissipation parameter $\Gamma$ is introduced to account for possible anomalous broadening of the conductance, which is often observed experimentally~\cite{DasSarma2016How,Liu2017Role}. Finite temperature acts as an additional (thermal) broadening mechanism -- the electron temperature may be well above the base temperature in experiments. We note that both $\Gamma$ and $V_c$ are nonessential aspects of our theory with respect to the ABS/MBS distinction-- we have them in the theory to make the results realistic, not because they are necessary for the main point of left/right ZBP-correlations being made in this work.

The potential profile for the device in Fig.\ref{fig:scheme} is assumed to contain a quantum dot at the left end that can generate low energy subgap Andreev bound states even in the non-topological phase \cite{Liu2017Andreev}.
We model the Hamiltonian of the quantum dot to be 
\begin{equation}\label{E3}
    \begin{aligned}
    H_{QD}=&\left(-\frac{\hbar^2}{2m^*}\partial_x^2-i\alpha_R\partial_x\sigma_y-\mu+V_{dot}(x)\right)\tau_z\\
    &+V_z\sigma_x-i\Gamma,
    \end{aligned}
\end{equation}
where $V_{dot}(x)=V_D\cos(3\pi x/2 l_D)$ is the confinement potential in the quantum dot. The quantum dot length $l_D$ is only a fraction of the total nanowire length $L$. The precise form of the quantum dot potential is irrelevant for our consideration as no qualitative conclusion depends on these details. A more thorough discussion of the model can be found in Ref.~\cite{Liu2017Andreev}.

The leads in the set-up in Fig.\ref{fig:scheme} are described by the BdG Hamiltonians 
\begin{equation}\label{E4}
    \begin{aligned}
    H_{lead}=&\left(-\frac{\hbar^2}{2m^*}\partial_x^2-i\alpha_R\partial_x\sigma_y-\mu+E_{lead}\right)\tau_z\\
    &+V_z\sigma_x-i\Gamma,
    \end{aligned}
\end{equation}
where an additional on-site energy $E_{lead}$ is added as a gate voltage. Each lead induces a NS tunnel barrier at the junction connected to the nanowire. The tunnel barrier that controls the conductance into the Majorana nanowire is described by a BdG Hamiltonian:
\begin{equation}\label{E5}
    \begin{aligned}
    H_{barrier}=&\left(-\frac{\hbar^2}{2m^*}\partial_x^2-i\alpha_R\partial_x\sigma_y-\mu+V_{barrier}(x)\right)\tau_z\\
    &+V_z\sigma_x-i\Gamma,
    \end{aligned}
\end{equation}
where $V_{barrier}=E_{barrier}\Pi_{l_{barrier}}(x)$ is a box-like potential with height $E_{barrier}$ and width $l_{barrier}$.

Note that the Hamiltonian in Eq.\eqref{E1}-\eqref{E5} do not overlap with each other in real space in our calculations. From the left most end, it starts with $H_{lead}$, and then $H_{barrier}$, $H_{QD}$, $H_{NW}$, $H_{barrier}$ and $H_{lead}$ to the right, as they are setup in Fig.\ref{fig:scheme}. Other than the infinitesimal dissipation term $i\Gamma$, mostly from the vortices in the parent superconductor, that may decrease the conductance \cite{DasSarma2016How,Liu2017Role}, we also take into account the effect of temperature, which also smears the conductance profiles. The finite-temperature conductance $G_T$ is then calculated from the zero-temperature conductance $G_0$ by the convolution of the derivative of the Fermi-Dirac distribution, i.e.,
\begin{equation}\label{E6}
	G_T(V)=-\int_{-\infty}^{\infty}dE G_0(E)\frac{df(E-V)}{dE}
\end{equation}
We numerically calculate the zero-temperature conductance $G_0=dI/dV$ by discretizing the Hamiltonian (Eq.\eqref{E1}-\eqref{E5}) into a lattice chain of tight-binding model\cite{Liu2017Andreev}, and obtaining the scattering matrix\cite{Setiawan2015Conductance} from the Python package KWANT \cite{Groth2014Kwant} for quantum transport. The zero-temperature conductance (in unit of $e^2/h$) is computed using the following formula.
\begin{equation}\label{E7}
	G_0=2+\sum_{\sigma,\sigma'=\uparrow,\downarrow}
	\left(|r_{eh}^{\sigma\sigma'}|^2-|r_{ee}^{\sigma\sigma'}|^2\right),
\end{equation}
where $r_{eh}$ and $r_{ee}$ are the Andreev and normal reflection amplitudes, respectively. The factor of 2 in Eq.\eqref{E7} is contributed by the two spin channels while we consider a one-subband system. The generalization to multisubband situations is straightforward, but adds no new element to the left/right tunneling correlations being discussed in this work. Since there is no experimental information on how many subbands are occupied in realistic situation, we stick to the minimal model of single orbital subband in the nanowire since the physics of interest here is completely independent of subband occupancy.

The conductance in the tunneling limit can be understood from the local density of states that can be estimated from the energy spectrum $E$ and the wave-function density $|\Psi(x)|^2$. To calculate these, we ignore the tunnel barrier effect from the leads, considering only the semiconductor nanowire coupled with the quantum dot. The Hamiltonian is a combination of Eq.\eqref{E1} and Eq.\eqref{E3}.
\begin{equation}\label{E8}
	\begin{aligned}
	&H_{tot}=H_{QD}+H_{NW}+H_t,\\
	&H_t=u+u^\dagger=\hat{f}_\alpha^\dagger\left(-t\delta_{\alpha\beta}+i\alpha_R\sigma_{\alpha\beta}^y\right)\hat{c}_\beta+h.c.,
	\end{aligned}
\end{equation}
where $H_{QD}$ is the isolated quantum dot, $H_{NW}$ is the semiconductor nanowire and $H_t$ is the coupling between them. $\hat{f}$ creates an electron at the end of the dot adjacent to the nanowire, while $\hat{c}$ annihilates an electron at the end of the nanowire connected to the dot. Then we diagonalize the total Hamiltonian in Eq.\eqref{E8}, obtaining the spectrum shown in Fig.\ref{fig:noSelf2} and Fig.\ref{fig:noSelf9}.

Our goal in this paper is to see the (non)correlations of the ZBCPs probed from leads on both ends when the MBS (ABS) appear. We not only analyze the conductance, but also carefully look into the form of the lowest lying wave functions and the nanowire energy spectra. Thus, it is important for us to separate the eigenstates of Eq.\eqref{E8} and combine them into the form of Majorana modes. Here we follow the recipe of Ref. \cite{Stanescu2018Illustrated} to decompose the finite energy Andreev states into Majorana. Consider a low-energy eigenfunction $\phi_\epsilon$ of a positive energy $\epsilon\ll\Delta$ such that this eigenfunction at position $n$ is represented as $\phi_\epsilon(n)=\left(u_{n\uparrow},u_{n\downarrow},v_{n\uparrow},v_{n\downarrow}\right)^T$ in Nambu space. Particle-hole symmetry guarantees the existence of a eigenfunction of negative energy $-\epsilon$ described by $\phi_{-\epsilon}(n)=(v_{n\uparrow}^*,v_{n\downarrow}^*,u_{n\uparrow}^*,u_{n\downarrow}^*)^T$. Combining these eigenstates, we construct the states of the form satisfying the Majorana conditions in Eq.\eqref{E9}.
\begin{subequations}\label{E9}
	\begin{align}
		&\psi_A(n)=\frac{1}{\sqrt{2}}\left[\phi_\epsilon(n)+\phi_{-\epsilon}(n)\right]\\
		&\psi_B(n)=-\frac{i}{\sqrt{2}}\left[\phi_\epsilon(n)-\phi_{-\epsilon}(n)\right]
	\end{align}
\end{subequations}
Here, $\psi_{\alpha}(n)=(\tilde{u}_{\alpha n\uparrow},\tilde{u}_{\alpha n\downarrow},\tilde{u}_{\alpha n\uparrow}^*,\tilde{u}_{\alpha n\downarrow}^*)^T$ have a spinor structure, where $\alpha=A,B$. Besides, $\tilde{u}_{A,n,\sigma}=u_{n\sigma}+v_{n\sigma}^*$ and $\tilde{u}_{B,n,\sigma}=-i(u_{n\sigma}-v_{n\sigma}^*)$, which meet the Majorana condition. Generally, $\psi_A$ and $\psi_B$ are not eigenstates of the BdG Hamiltonian, except for $\epsilon=0$, while the Majorana representation of the eigenstates of the BdG Hamiltonian, $\phi_{\pm\epsilon}=\frac{1}{\sqrt{2}}(\psi_A\pm i\psi_B)$ is generic. In general, one can worry about a phase ambiguity in the wave-functions. However, the model considered here (Eq.\eqref{E1}) has a chiral symmetry (i.e. anti-commutes with $\sigma_y\tau_y$), so that we can choose $\phi_{-\epsilon}=\sigma_y\tau_y\phi_\epsilon$. This eliminates the phase ambiguity problem.

In our calculations, we obtain the Majorana-form wave function probabilities $|\psi_A|^2$ and $|\psi_B|^2$ over the spatial space in the nanowire, by first summing over the inner degrees of freedom (spin, particle-hole spaces). If $|\psi_A|^2$ and $|\psi_B|^2$ are localized separately at two ends of nanowire, then they are an MBS pair, which means the ZBCP we see in the corresponding conductance plot results from Majorana zero modes. On the contrary, if we cannot separate them clearly by Eq.\eqref{E9}, then the ZBCP comes from ABS. By comparing the behavior of wavefunctions on both ends, we can distinguish the sources of ZBCP. We emphasize that the trivial almost-zero-energy ABS here are all composed of double MBS modes which overlap strongly spatially. When the two MBS are well-separated without overlap, being localized at the two ends of the wire, we have topological Majorana zero modes (provided, of course, that the nanowire is long enough).

Considering that the theoretical methodology in our current work is standard (although the basic questions we ask and answers we provide are new) and has been widely studied in the literature \cite{Sau2010NonAbelian,Stanescu2011Majorana,Lin2012Zerobias,Setiawan2015Conductance,DasSarma2016How,Liu2017Role,Liu2017Andreev,Stanescu2018Illustrated}, we do not provide any further details about the theory. Instead, we focus on the numerical results we compute based on above methods.

The parameters in all of our numerical results (Figs.\ref{fig:noSelf2}, \ref{fig:noSelf9}, \ref{fig:CrossG}, \ref{fig:noSelf1}-\ref{fig:GSelfE}) are chosen as follows (with the InSb nanowires in mind, although we do not attempt any quantitative comparison with experiments because of the large number of unknown parameters in the semiconductor-superconductor hybrid system, e.g., the spin-orbit coupling, the effective mass, the effective g-factor, the lead-nanowire tunnel coupling, the superconductor-nanowire tunnel coupling, the chemical potential, the active wire length, the applicable coherence length in the nanowire, quantum dot confinement potential). The effective mass is $m^*=0.015m_e$, nanowire length $L=5$ $\micro m$, induced superconducting gap $\Delta_0=0.9$ meV, spin-orbit coupling $\alpha_R=0.5$ eV$\AA$. The gate voltage in the lead is $E_{lead}=-25$ meV, with the induced tunnel barrier height $E_{barrier}=10$ meV, and the barrier length $l_{barrier}=20$ nm. The strength of the confinement potential in the quantum dot is $V_D=4$ meV, with length $l_D=0.3$ $\micro m$. The temperature, which smears the conductance profile by thermally broadening all sharp features, is set at 0.02 meV. The phenomenological dissipation parameter is $\Gamma=0.01$ meV. The above parameters will be fixed throughout for all the cases in our results, and other tuning parameters, including the chemical potential $\mu$, the Zeeman energy $V_z$,  along with the superconducting gap collapsing point $V_c$, will be provided in the captions of the figures. The topological quantum phase transition (TQPT) field
\begin{equation}\label{E10}
    V_{Z_c}=\sqrt{\mu^2+\Delta_0^2}
\end{equation}
is also provided in each case. We note that we only consider the case where $V_c>V_{Z_c}$, so in principle, the topological regime exists for $V_{Z_c}<V_z<V_c$ so that MBS-induced ZBCPs can manifest itself.  Experimentally, the situation $V_z<V_{Z_c}$ is allowed, and in such a case, all ZBCP must arise from trivial MBS.

\section{Results and Discussions}\label{sec:level1_3}
The key idea underlying the current work is simple:  Topological Majorana modes have nonlocal correlations, and any zero bias peak associated with MBS must manifest itself in tunneling from each end of the wire, since MBS must always exist in pairs at both ends, whereas by contrast, ABS is a non-topological subgap fermionic bound state which will be randomly localized near one or the other end of the wire and as such will have no nonlocal correlations.  The implication of this simple idea is that ABS-induced ZBCP would arise only when tunneling connects to the relevant ABS, which is necessarily at one end, thus ABS-induced ZBCP are not correlated from both ends, whereas MBS-induced ZBCP must necessarily be correlated.  The details of this implication are, however, quite subtle (and depend crucially on system parameters, particularly, the chemical potential) as the results of our extensive numerical simulations of nanowire tunneling conductance show.  We discuss these results below.

\begin{figure}[htbp]
	\includegraphics[scale=0.16]{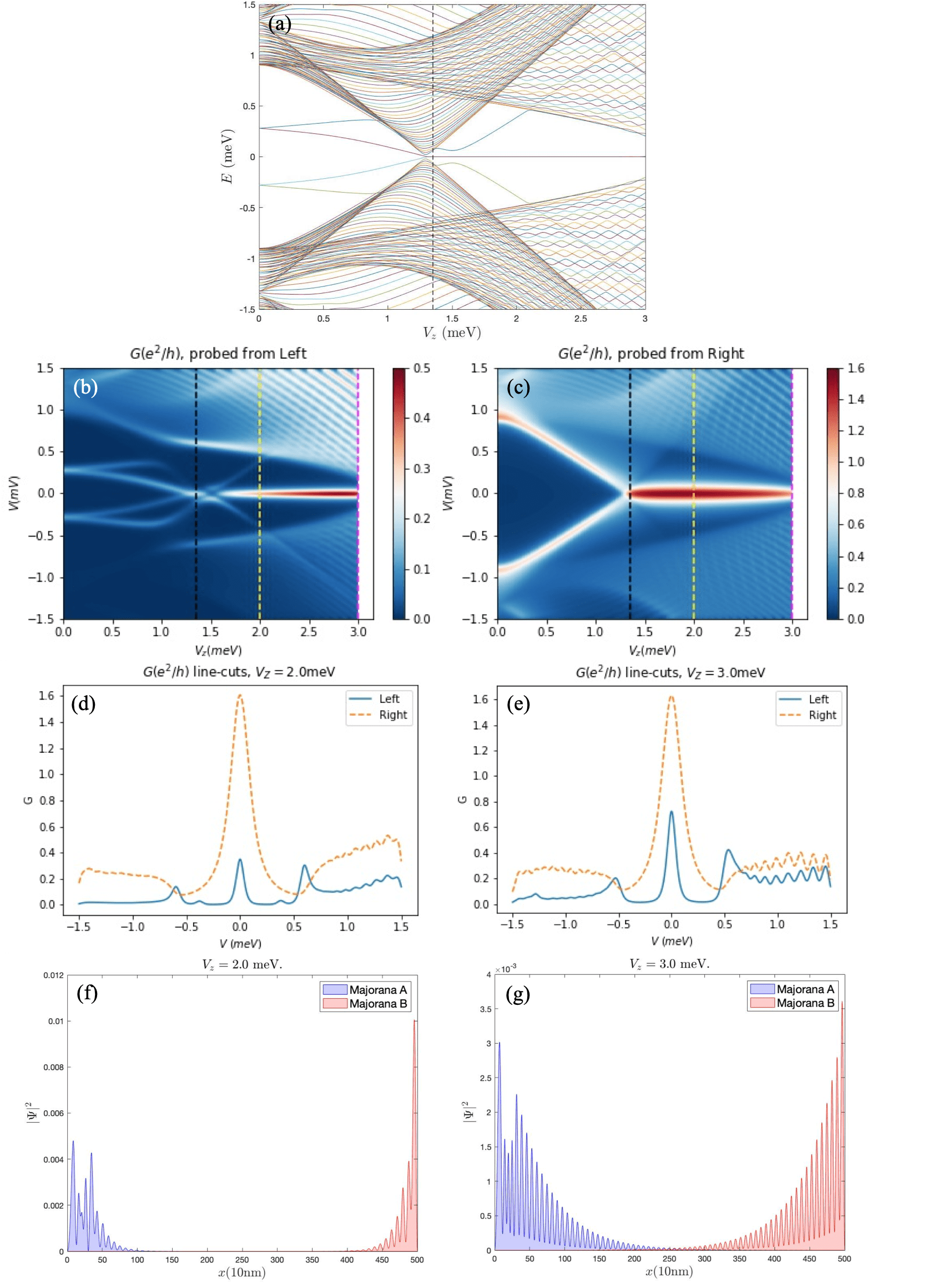}
	\caption{(a) Energy spectrum. (b) conductance $G(V)$ measured from the left lead. (c) conductance $G(V)$ measured from the right lead. (d) conductance line cut at $V_z=2.0$ meV. (e) conductance line cut at $V_z=3.0$ meV. (f) Majorana components of lowest lying wave functions in the nanowire at $V_z=2.0$ meV, above $V_{Z_c}$. (g) Majorana components of lowest lying wave functions in the nanowire at $V_z=3.0$ meV, above $V_{Z_c}$. The superconducting collapsing field is at $V_c=3.2$ meV. The black, yellow and purple vertical dashed lines in the panels (a,b,c) represent the Zeeman strengths $V_{Z_c}=1.35$ meV, $V_z=2.0$ meV and 3.0 meV respectively. The chemical potential is chosen to be $\mu=1.0$ meV.}
	\label{fig:noSelf2}
\end{figure}

The spectrum of the nanowire shown in Fig.\ref{fig:scheme} in the small chemical potential regime ($\mu=1$ meV) is shown in Fig.\ref{fig:noSelf2}(a). In addition to the zero bias Majorana state that appears above $V_{Z_c}=1.35 meV$ the spectrum in Fig.\ref{fig:noSelf2}(a) shows a pair of ABS states that approach and stick to near zero energy near $V_{Z_c}$. However, the ABS stays near zero energy only for a small range of Zeeman potential. The low-energy states in the spectrum appear from sub-gap bound states that contribute to the conductance spectra from the tunneling into the left and right end of the nanowire shown in Figs.\ref{fig:noSelf2}(b,c) respectively. It is clear from Fig.\ref{fig:noSelf2}(b), the wave-functions of the ABS states that arise from the quantum dot on the left in Fig.\ref{fig:scheme} are localized near the quantum dot and lead to features only in the conductance from the left end. In contrast, the conductance from the right end (Fig.\ref{fig:noSelf2}(c)) shows only a zero energy bound state above $V_{Z_c}$. Despite the conductance profiles in Figs.\ref{fig:noSelf2}(b,c) looking quite different, the zero-bias peaks on both the left and the right appear at the same Zeeman potential ($V_{Z_c}$). This is because both zero bias peaks arise from MZMs and the ABSs in Fig.\ref{fig:noSelf2}(a) do not stick to zero energy. An examination of the line-cuts of the conductance in Fig.\ref{fig:noSelf2}(b) near the onset shown in Fig.\ref{fig:noSelf2}(d) demonstrate that despite the correlated onset of zero-bias conductance, the height of the peaks at the left and the right end are quite different. However, the conductances from the left and right end become comparable (both approaching quantization) (Fig.\ref{fig:noSelf2}(e)) for Zeeman potentials far above the onset. For this regime of low chemical potential, the zero-energy state arises from Majorana modes appearing at each of the ends of the wire. This is confirmed from an examination of the wave-functions of the near zero energy states at both nearer to the onset of the TQPT field ($V_{Z_c}$)  (Fig.\ref{fig:noSelf2}(f)) and deeper in the topological phase (Fig.\ref{fig:noSelf2}(g)). Both of these plots show finite weight for the wave-function amplitude at either end.

\begin{figure}[htbp]
	\includegraphics[scale=0.16]{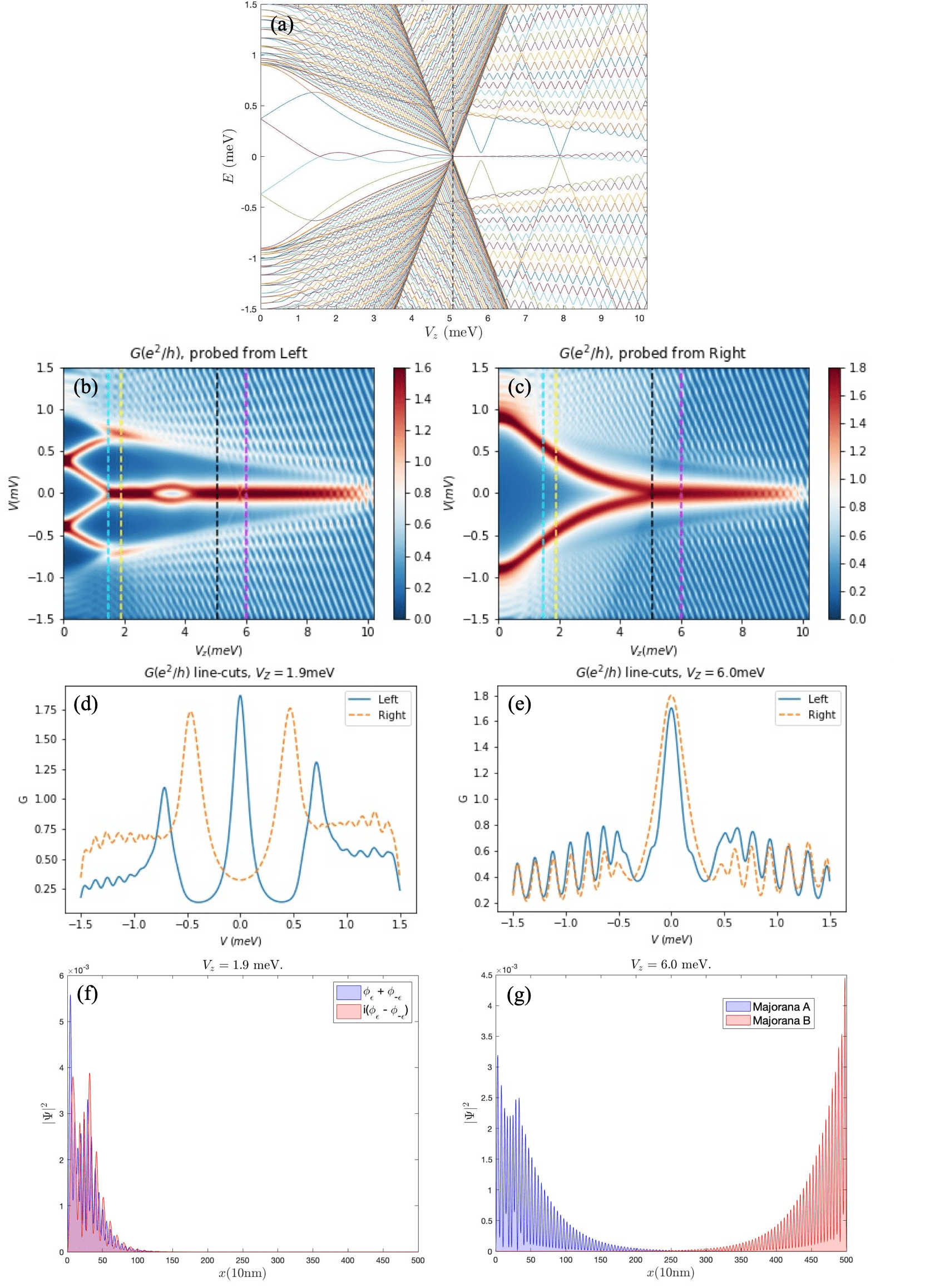}
	\caption{(a) Energy spectrum. (b) conductance $G(V)$ measured from the left lead. (c) conductance $G(V)$ measured from the right lead. (d) conductance line cut at $V_z=1.9$ meV. (e) conductance line cut at $V_z=6.0$ meV. (f) Majorana components of lowest lying wave functions in the nanowire at $V_z=1.9$ meV, below $V_{Z_c}$. (g) Majorana components of lowest lying wave functions in the nanowire at $V_z=6.0$ meV, above $V_{Z_c}$. The superconducting collapsing field is at $V_c=10.4$ meV. The black, yellow and purple vertical dashed lines in the panels (a,b,c) represent the Zeeman strengths $V_{Z_c}=5.08$ meV, $V_z=1.9$ meV and 6.0 meV respectively. The chemical potential is chosen to be $\mu=5.0$ meV.}
	\label{fig:noSelf9}
\end{figure}

Raising the chemical potential $\mu$ to 5 meV changes the situation qualitatively. As seen in Fig.\ref{fig:noSelf9}(a), now ABSs appear in the spectrum which stick to zero energy even for $V_z<V_{Z_c}$\cite{Liu2017Andreev,Stanescu2018Illustrated,Moore2018Quantized}. In this case we see that a zero-bias peak appears on the left (Fig.\ref{fig:noSelf9}(b)) at a substantially lower Zeeman potential below $V_{Z_c}$, where the zero-bias peak appears on the right (Fig.\ref{fig:noSelf9}(c)). The line cuts of the conductance plots at the onset of the ZBCP on the left (Fig.\ref{fig:noSelf9}(d))  shows a dramatic difference between the left and the right end. The former has a ZBCP and the latter has strong gap features. As the Zeeman field is increased, the ZBCPs at both ends become similar (Fig.\ref{fig:noSelf9}(e)) as expected in the topological phase (Fig.\ref{fig:noSelf2}(e)). The non-topological origin for the ZBCP in Fig.\ref{fig:noSelf9}(d) becomes clear from an examination of the wave-functions at the corresponding Zeeman field. We see from Fig.\ref{fig:noSelf9}(d), that the wave-function is entirely located at one end as expected from ABS. Decomposing the ABS into a pair of Majorana states (using Eq.\eqref{E9}), we see that both Majorana components of the ABS are located at the same end. This is in contrast to higher Zeeman field (Fig.\ref{fig:noSelf9}(e)), where the Majoranas move to opposite end of the wires as expected for a true topological state. However, the conductance in this topological state is essentially identical to the non-topological one from the left end in Fig.\ref{fig:noSelf9}(d).

In the main text, we only show results for two chemical potentials in Figs.\ref{fig:noSelf2} and \ref{fig:noSelf9} to demonstrate the fundamental nature of correlated and uncorrelated end-to-end conductances from the two wire ends for MBS and ABS respectively.  In order to establish the generic nature of such correlations for MBS (or lack thereof for ABS), we present extensive results in the Appendix for many other values of chemical potential and magnetic field, clearly showing that the presence or absence of end-to-end conductance correlations implies the presence or absence of MBS or ABS respectively in the nanowire.

\begin{figure}[t]
	\includegraphics[scale=0.15]{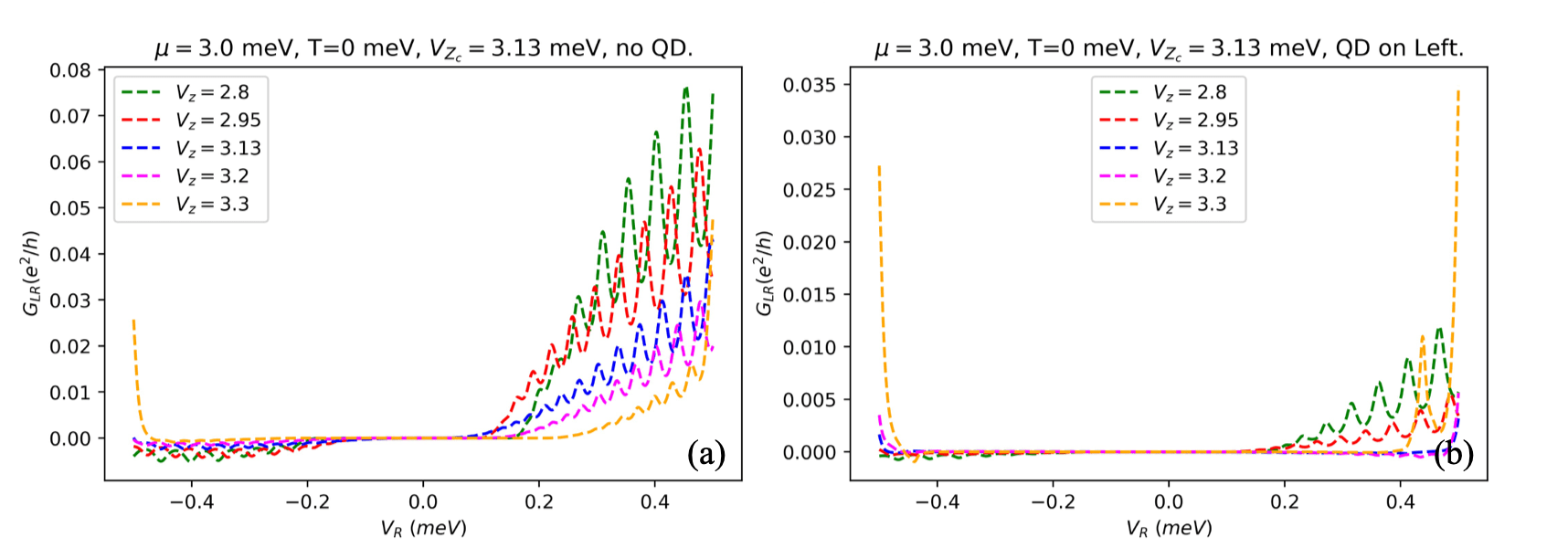}
	\caption{Cross-conductance $G_{LR}$ as a function of applied voltage $V_R$ for different Zeeman splittings $V_Z$ applied to the Majorana wire. Panels (a) and (b) show $G_{LR}$ for an ideal Majorana wire and one with a quantum dot on the left respectively. The cross-conductance vanishes below the bulk gap. The bulk gap in (a) appears to close and reopen as $V_Z$ crosses the critical point. Such a gap closing is difficult to see in (b). Note that the induced tunnel barrier height is $E_{barrier}=1$ meV.}
	\label{fig:CrossG}
\end{figure}

In addition to conductance correlation between the two ends, the three terminal set-up in Fig.\ref{fig:scheme} allows access to two cross conductances $G_{LR}=dI_L/dV_R$ and $G_{RL}=dI_R/dV_L$. Such a non-local conductance has been proposed as a way to distinguish bulk states from potential inhomogeneity-induced sub-gap states \cite{Rosdahl2018Andreev}. The appearance of bulk states at the gap closure of the Majorana nanowire is a hall-mark of the TQPT at which Majorana modes are supposed to appear. Such a TQPT is thus characterized by a signature associated with bulk gap closure as seen in the cross-conductance $G_{LR}$ versus voltage (Fig.\ref{fig:CrossG}(a)). While, as a matter of principle, the cross-conductance reveals the bulk gap closure, the cross-conductance is constrained to vanish at zero-voltage by particle-hole symmetry, which has motivated 
other non-local signatures such as heat transport \cite{Akhmerov2011Quantized} and non-linear conductivity \cite{Fregoso2013Electrical}. Because of this, the gap-closure at the critical Zeeman field in Fig.\ref{fig:CrossG}(a) appears more as a soft-gap where the conductance continues to vanish at zero voltage. In Fig.\ref{fig:CrossG}(b), we plot the cross-conductance as a function of voltage for the case with a quantum dot on the left. We find that the quantum dot strongly suppresses the bulk gap closing signature near the critical Zeeman field by reducing the conductance scale further near zero voltage. Therefore, the observation of gap closing in cross-conductance might require appropriate tuning/engineering of the end potential to eliminate the effect of the quantum dots on the cross-conductance. This can probably be done with additional gate voltages to tune system parameters appropriately.

\section{Conclusion}\label{sec:level1_4}
We have shown by extensive numerical simulations that the zero bias tunneling conductance peaks arising from trivial ABS can be distinguished from those arising from topological MBS, by comparing separate tunneling measurements carried out simultaneously from the two ends of the wire. The fact that the MBS-induced ZBCPs are nonlocally correlated in tunneling measurements from the two ends was already emphasized in Ref.~\cite{DasSarma2012Splitting}, but the current work incorporates the complications arising from the existence of nontopological ABS in the wire, a situation not addressed in Ref.~\cite{DasSarma2012Splitting}. The basic idea behind our work is simple: Topological MBS are intrinsic nonlocal objects with their wavefunctions existing at both ends of the wire, whereas the nontopological ABS can only exist near one or the other end depending on the extrinsic details giving rise to the ABS. Thus, an NS tunneling measurement from a particular wire end can only probe an ABS if it exists near that end, whereas the MBS, if it exists, must be equally accessible from both ends because of its nonlocal nature. This simple physics is reflected in our simulations and should be observable experimentally. 

One experimental (or materials) issue may severely compromise the observation (and verification) of MBS in realistic nanowires which we now discuss. This is the issue of the experimentally observed collapse of the parent bulk gap at some magnetic field$\sim1T$ (represented by $V_c$ in our simulations) universally happening in all the Majorana nanowire experiments. The reason for this gap collapse is unclear, and one possibility is a strong magnetic field induced orbital effect in the parent superconductor. It could also be arising simply from the parent SC reaching the Clogston limit with the Zeeman spin splitting becoming equal to the bulk SC gap. Denoting the corresponding Zeeman splitting for this gap collapse by $V_c$ (to be contrasted with the critical TQPT field $V_{Z_c}$ defining the topological regime $V_z>V_{Z_c}$), it is obvious that if $V_c<V_{Z_c}$, MBS cannot appear in the system as an isolated anyon. Unfortunately, $V_{Z_c}$ is unknown in experiments, only $V_c$ is known. For clarity, we have limited the simulations in this work to the case $V_{Z_c} < V_c$. However, since the superconductor self-energy\cite{Stanescu2010Proximity} becomes purely imaginary and number conserving for $V_z>V_c$, we do not expect any bound states in this regime. This is consistent with experimental measurements, where typically, nothing happens experimentally for $V_z>V_c$ with all conductance signals in the nanowire basically vanishing above this parent SC gap collapse point. If $V_{Z_c}>V_c$ generically (e.g. perhaps because the chemical potential is always rather large, making $V_{Z_c}$ also large), then all observed ZBCPs arise from ABS and MBS are simply inaccessible until $V_{Z_c}<V_c$ can be achieved by tuning system parameters. Obviously, any topological MBS can exist only in the Zeeman field regime $V_{Z_c}<V_z<V_c$. Therefore, if $V_c<V_{Z_c}$ (with the SC bulk gap collapsing before $V_z$ reaches the TQPT point), our proposed nonlocal correlation experiments would not work, and all observed ZBCPs would manifest no correlations in the tunneling data sets from the two wire ends, since they all must be ABS in such an unfortunate scenario. Making $V_c>V_{Z_c}$ is an all-important materials challenge deserving serious experimental efforts for achieving further progress in the search for topological Majorana modes in superconductor-semiconductor hybrid systems.

One point to note here is that the nanowire should be 'long' (i.e. longer than the superconducting coherence length) for topological physics to manifest itself. Thus, our results and conclusions about the presence or absence of end-to-end conductance correlations implying the presence or absence respectively of MBS and ABS strictly applies to long wires. In short wires, the ABS wavefunctions at the two ends may happen to overlap, depending on the details, leading to apparent conductance correlations even for trivial Andreev bound states. In such a short-wire situation, end-to-end conductance correlations would not be able to distinguish between MBS and ABS. This is, however, a rather trivial point since topological Majorana modes cannot exist in short wires any way since overlap between the MBS wavefunctions from the two ends would lift the topological degeneracy coupling of the two MBS. In short wires, it is easier to discern MBS by studying the MBS oscillations directly as emphasized in Ref.\cite{DasSarma2012Splitting} already.  Our current work applies to long wires where there is a fundamental distinction between ABS and MBS in contrast to short wires, where there is no essential difference between the two because of strong wavefunction overlap between the two ends.

Another potential false positive possibility for our proposed end-to-end conductance correlations as a test for the MZM existence is that it is possible, in principle, for two different accidental (and therefore, unknown) quantum dots to be present at the two ends of the wire leading to two Andreev bound states at two ends providing ZBCP correlations in tunneling measurements from the two ends. This problem is probabilistically less likely, but cannot be ruled out. Therefore, much care is necessary in concluding about the MZM existence just from correlation measurements of ZBCP-- it is desirable also to have other measurements showing the closing/opening of a bulk gap at the TQPT precisely where the ZBCP starts forming in the tunneling measurements.

We believe that the end-to-end tunneling correlation measurements can be carried out in the laboratory right now, and are encouraged by the fact that several groups are currently trying to implement such multiprobe tunneling measurements where NS tunneling probes are used in several contacts along the wire \cite{Grivnin2019Concomitant,Kouwenhovenprivate,Marcusprivate,Manfraprivate}. In particular, Grivnin et al. \cite{Grivnin2019Concomitant} has already shown the way by carrying out a pioneering mulitprobe tunneling measurement searching for the simultaneous closing and opening of a bulk gap along with the appearance of a zero bias peak \cite{Huang2018Quasiparticle,Stanescu2012Close}. It should be straightforward to adapt this multiprobe set up to carry out simultaneous NS tunneling measurements from both ends of the wire to look for conductance correlations as proposed in our work. One interesting conclusion of Grivnin et al. \cite{Grivnin2019Concomitant} is that there are different kinds of conductance zero bias peaks in the nanowires, and not all zero bias peaks are similar. This is of course the precise conclusion of the current work also (ABS-induced and MBS-induced zero bias peaks are fundamentally different with respect to nonlocal correlations), but much more work along the line of correlated tunneling spectroscopy from both ends is necessary before any firm conclusion is possible. In particular, tunnel probes themselves may introduce ABS and hence ABS-induced ZBCPs, complicating the experimental situation \cite{Stanescu2018Building}, but our conclusion about nonlocal correlations in the MBS-induced ZBCP in contrast to lack of correlations in ABS-induced ZBCPs would still apply.

For completeness, we have also studied the cross-conductance that can be measured in the same multi-terminal set-up as the conductance correlation. Such measurements have been proposed as a way to detect the TQPT, which might help separate ABSs and MBSs \cite{Rosdahl2018Andreev}. However, similar to conductance correlations, the presence of end quantum dot induced ABSs might obscure the signature of the TQPT. Therefore, the failure to observe the TQPT (and the bulk gap closure) in the cross-correlation would not necessarily imply the absence of MBS. One could likely combine information from cross-conductance and conductance correlations to find MBSs where both tests give a weak signal. Similarly, the vanishing of cross-conductance in the Majorana wire away from zero energy can be used to estimate the coherence length relative to the length of the wire. One serious problem is that current experiments provide no information on either the magnitude of the applicable coherence length or the location of the TQPT (i.e. the value of $V_{Z_c}$) since the bulk gap closing feature has not yet been directly observed experimentally.

In conclusion, we have shown that the presence (absence) of correlated zero bias conductance from the two ends as a function of the applied magnetic field could indicate the presence (absence) of topological Majorana (trivial Andreev) bound states in nanowires.

Acknowledgement: This work is supported by Microsoft and Laboratory for Physical Sciences. Yi-Hua Lai thanks Chunxiao Liu, Haining Pan, and Yingyi Huang for helpful discussions and suggestions, particularly on the use of KWANT in the numerical simulations. The authors acknowledge the support of the University of Maryland High Performance Computing Center for the use of Deep Through II cluster for carrying out the numerical work. 

\bibliography{BibNonLocalCorr.bib}

\clearpage
\onecolumngrid
\vspace{1cm}
\begin{center}
	{\bf\large Appendix}
\end{center}
\vspace{0.5cm}
\setcounter{secnumdepth}{3}
\setcounter{equation}{0}
\setcounter{figure}{0}
\renewcommand{\thefigure}{A\arabic{figure}}
\newcommand\Scite[1]{[S\citealp{#1}]}
\makeatletter \renewcommand\@biblabel[1]{[S#1]} \makeatother

\begin{figure*}[hbt]
	\includegraphics[scale=0.2]{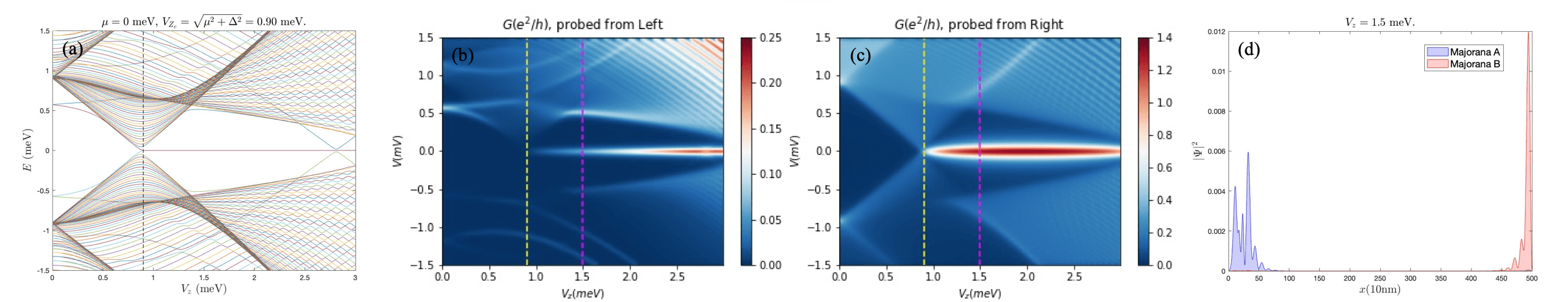}
	\caption{Energy spectrum, conductance measured from both ends, and the lowest lying wave functions above $V_{Z_c}=0.90$ meV, for the case of chemical potential $\mu=0$ meV. (a): energy spectrum. (b): conductance $G(V)$ measured from the left lead. (c): conductance $G(V)$ measured from the right lead. (d): lowest lying wave functions in the nanowire at $V_z=1.5$ meV, above $V_{Z_c}$. The superconducting collapsing field is at $V_c=3.2$ meV. The black dashed line in (a) and yellow dashed lines in (b,c) are at $V_{Z_c}=0.90$ meV; the purple dashed lines in (b,c) are at $V_z=1.5$ meV.}
	\label{fig:noSelf1}
\end{figure*}

\begin{figure*}[hbt]
	\includegraphics[scale=0.17]{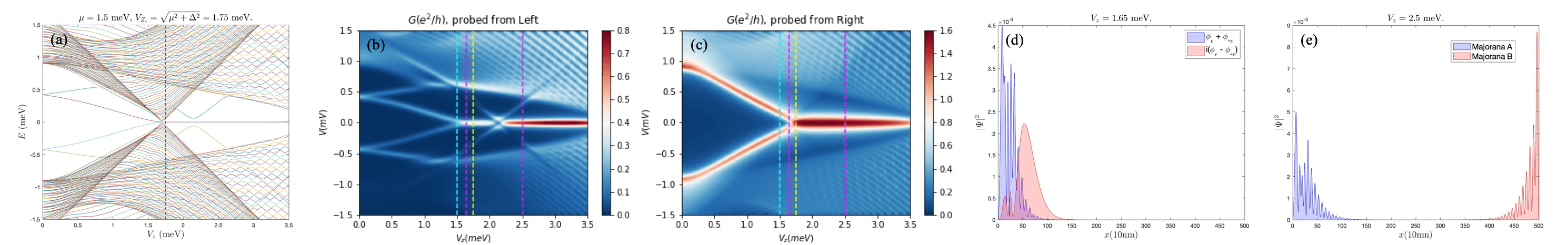}
	\caption{Energy spectrum, conductance measured from both ends, and the lowest lying wave functions below and above $V_{Z_c}=1.75$ meV, for the case of chemical potential $\mu=1.5$ meV. (a): energy spectrum. (b): conductance $G(V)$ measured from the left lead. (c): conductance $G(V)$ measured from the right lead. (d): lowest lying wave functions in the nanowire at $V_z=1.65$ meV, below $V_{Z_c}$. (e): lowest lying wave functions in the nanowire at $V_z=2.5$ meV, above $V_{Z_c}$. The superconducting collapsing field is at $V_c=3.7$ meV. The black dashed line in (a) and yellow dashed lines in (b,c) are at $V_{Z_c}=1.75$ meV; the cyan dashed lines in (b,c) are at $V_z=1.5$ meV, where the ZBCP firstly starts; the first purple dashed lines in (b,c) are at $V_z=1.65$ meV; the second purple dashed lines in (b,c) are at $V_z=2.5$ meV.}
	\label{fig:noSelf3}
\end{figure*}

\begin{figure*}[hbt]
	\includegraphics[scale=0.17]{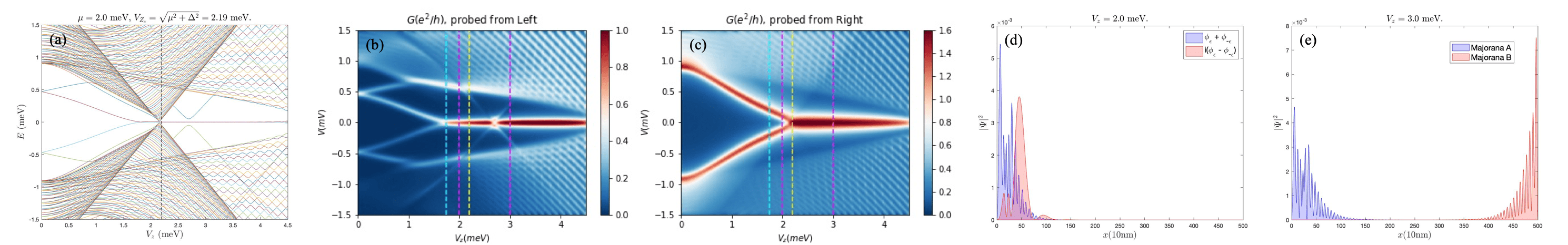}
	\caption{Energy spectrum, conductance measured from both ends, and the lowest lying wave functions below and above $V_{Z_c}=2.19$ meV, for the case of chemical potential $\mu=2.0$ meV. (a): energy spectrum. (b): conductance $G(V)$ measured from the left lead. (c): conductance $G(V)$ measured from the right lead. (d): lowest lying wave functions in the nanowire at $V_z=2.0$ meV, below $V_{Z_c}$. (e): lowest lying wave functions in the nanowire at $V_z=3.0$ meV, above $V_{Z_c}$. The superconducting collapsing field is at $V_c=4.7$ meV. The black dashed line in (a) and yellow dashed lines in (b,c) are at $V_{Z_c}=2.19$ meV; the cyan dashed lines in (b,c) are at $V_z=1.75$ meV, where the ZBCP firstly starts; the first purple dashed lines in (b,c) are at $V_z=2.0$ meV; the second purple dashed lines in (b,c) are at $V_z=3.0$ meV.}
	\label{fig:noSelf4}
\end{figure*}

\begin{figure*}[hbt]
	\includegraphics[scale=0.17]{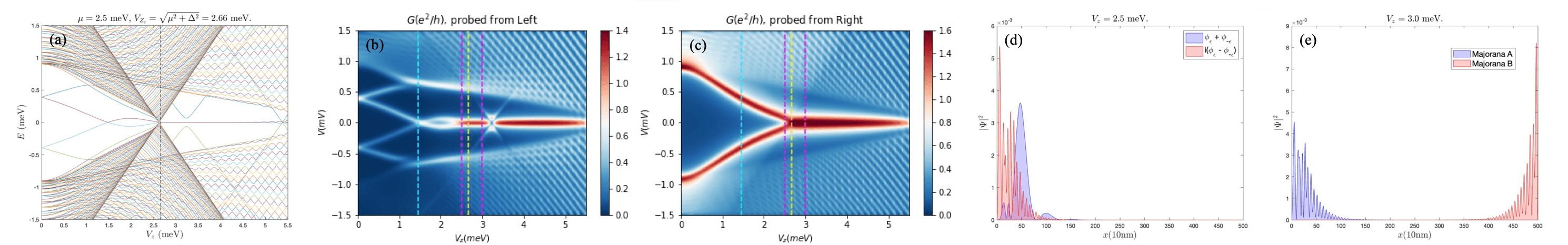}
	\caption{Energy spectrum, conductance measured from both ends, and the lowest lying wave functions below and above $V_{Z_c}=2.66$ meV, for the case of chemical potential $\mu=2.5$ meV. (a): energy spectrum. (b): conductance $G(V)$ measured from the left lead. (c): conductance $G(V)$ measured from the right lead. (d): lowest lying wave functions in the nanowire at $V_z=2.5$ meV, below $V_{Z_c}$. (e): lowest lying wave functions in the nanowire at $V_z=3.0$ meV, above $V_{Z_c}$. The superconducting collapsing field is at $V_c=5.7$ meV. The black dashed line in (a) and yellow dashed lines in (b,c) are at $V_{Z_c}=2.66$ meV; the cyan dashed lines in (b,c) are at $V_z=1.45$ meV, where the ZBCP firstly starts; the first purple dashed lines in (b,c) are at $V_z=2.5$ meV; the second purple dashed lines in (b,c) are at $V_z=3.0$ meV.}
	\label{fig:noSelf5}
\end{figure*}

\begin{figure*}[hbt]
	\includegraphics[scale=0.17]{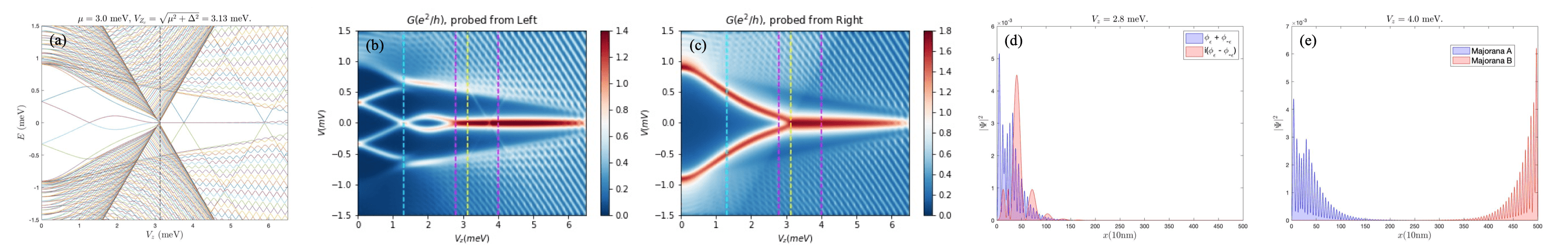}
	\caption{Energy spectrum, conductance measured from both ends, and the lowest lying wave functions below and above $V_{Z_c}=3.13$ meV, for the case of chemical potential $\mu=3.0$ meV. (a): energy spectrum. (b): conductance $G(V)$ measured from the left lead. (c): conductance $G(V)$ measured from the right lead. (d): lowest lying wave functions in the nanowire at $V_z=2.8$ meV, below $V_{Z_c}$. (e): lowest lying wave functions in the nanowire at $V_z=4.0$ meV, above $V_{Z_c}$. The superconducting collapsing field is at $V_c=6.7$ meV. The black dashed line in (a) and yellow dashed lines in (b,c) are at $V_{Z_c}=3.13$ meV; the cyan dashed lines in (b,c) are at $V_z=1.3$ meV, where the ZBCP firstly starts; the first purple dashed lines in (b,c) are at $V_z=2.8$ meV; the second purple dashed lines in (b,c) are at $V_z=4.0$ meV.}
	\label{fig:noSelf6}
\end{figure*}

\begin{figure*}[hbt]
	\includegraphics[scale=0.17]{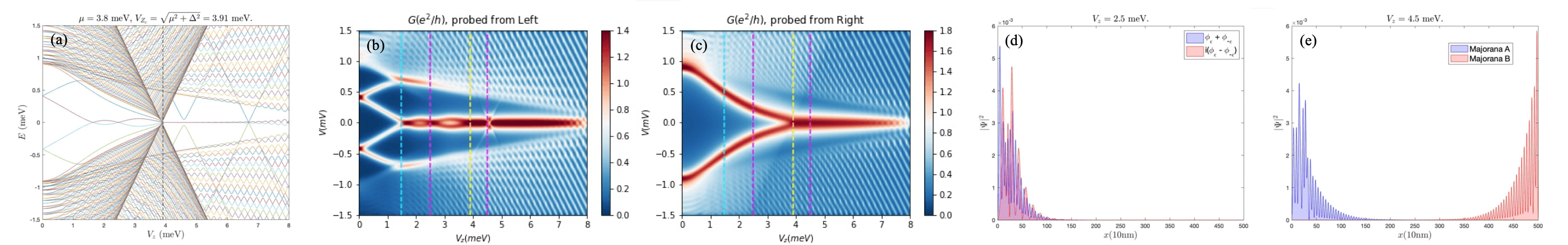}
	\caption{Energy spectrum, conductance measured from both ends, and the lowest lying wave functions below and above $V_{Z_c}=3.91$ meV, for the case of chemical potential $\mu=3.8$ meV. (a): energy spectrum. (b): conductance $G(V)$ measured from the left lead. (c): conductance $G(V)$ measured from the right lead. (d): lowest lying wave functions in the nanowire at $V_z=2.5$ meV, below $V_{Z_c}$. (e): lowest lying wave functions in the nanowire at $V_z=4.5$ meV, above $V_{Z_c}$. The superconducting collapsing field is at $V_c=8.2$ meV. The black dashed line in (a) and yellow dashed lines in (b,c) are at $V_{Z_c}=3.91$ meV; the cyan dashed lines in (b,c) are at $V_z=1.5$ meV, where the ZBCP firstly starts; the first purple dashed lines in (b,c) are at $V_z=2.5$ meV; the second purple dashed lines in (b,c) are at $V_z=4.5$ meV.}
	\label{fig:noSelf7}
\end{figure*}

\begin{figure*}[hbt]
	\includegraphics[scale=0.17]{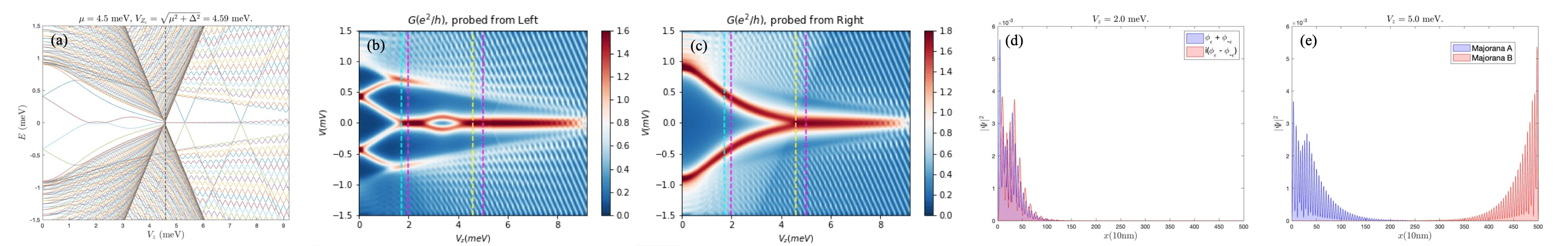}
	\caption{Energy spectrum, conductance measured from both ends, and the lowest lying wave functions below and above $V_{Z_c}=4.59$ meV, for the case of chemical potential $\mu=4.5$ meV. (a): energy spectrum. (b): conductance $G(V)$ measured from the left lead. (c): conductance $G(V)$ measured from the right lead. (d): lowest lying wave functions in the nanowire at $V_z=2.0$ meV, below $V_{Z_c}$. (e): lowest lying wave functions in the nanowire at $V_z=5.0$ meV, above $V_{Z_c}$. The superconducting collapsing field is at $V_c=9.4$ meV. The black dashed line in (a) and yellow dashed lines in (b,c) are at $V_{Z_c}=4.59$ meV; the cyan dashed lines in (b,c) are at $V_z=1.7$ meV, where the ZBCP firstly starts; the first purple dashed lines in (b,c) are at $V_z=2.0$ meV; the second purple dashed lines in (b,c) are at $V_z=5.0$ meV.}
	\label{fig:noSelf8}
\end{figure*}

\begin{figure*}[hbt]
	\includegraphics[scale=0.17]{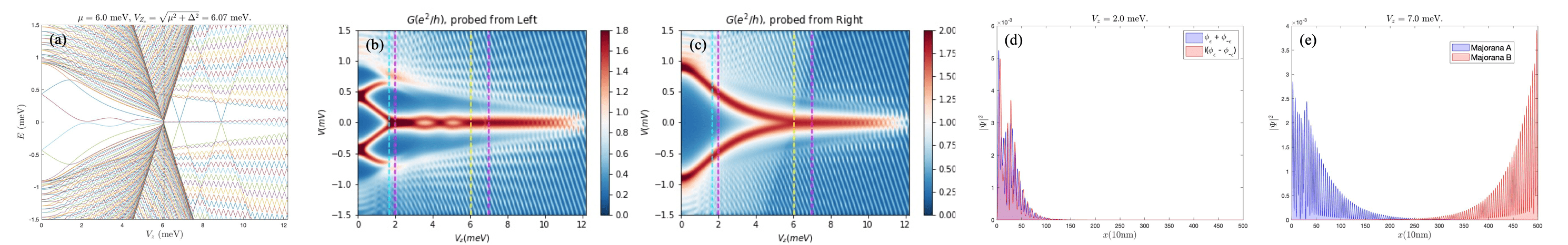}
	\caption{Energy spectrum, conductance measured from both ends, and the lowest lying wave functions below and above $V_{Z_c}=6.07$ meV, for the case of chemical potential $\mu=6.0$ meV. (a): energy spectrum. (b): conductance measured $G(V)$ from the left lead. (c): conductance $G(V)$ measured from the right lead. (d): lowest lying wave functions in the nanowire at $V_z=2.0$ meV, below $V_{Z_c}$. (e): lowest lying wave functions in the nanowire at $V_z=7.0$ meV, above $V_{Z_c}$. The superconducting collapsing field is at $V_c=12.4$ meV. The black dashed line in (a) and yellow dashed lines in (b,c) are at $V_{Z_c}=6.07$ meV; the cyan dashed lines in (b,c) are at $V_z=1.7$ meV, where the ZBCP firstly starts; the first purple dashed lines in (b,c) are at $V_z=2.0$ meV; the second purple dashed lines in (b,c) are at $V_z=7.0$ meV.}
	\label{fig:noSelf10}
\end{figure*}

\begin{figure*}[hbt]
	\includegraphics[scale=0.17]{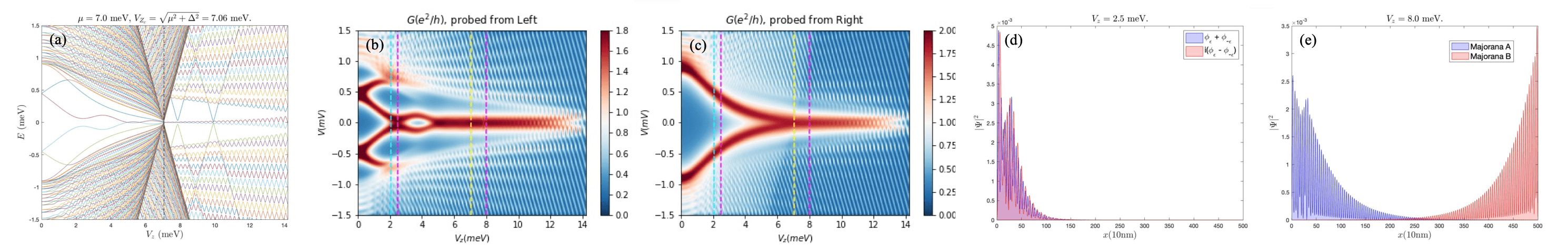}
	\caption{Energy spectrum, conductance measured from both ends, and the lowest lying wave functions below and above $V_{Z_c}=7.06$ meV, for the case of chemical potential $\mu=7.0$ meV. (a): energy spectrum. (b): conductance $G(V)$ measured from the left lead. (c): conductance $G(V)$ measured from the right lead. (d): lowest lying wave functions in the nanowire at $V_z=2.5$ meV , below $V_{Z_c}$. (e): lowest lying wave functions in the nanowire at $V_z=8.0$ meV, above $V_{Z_c}$. The superconducting collapsing field is at $V_c=14.4$ meV. The black dashed line in (a) and yellow dashed lines in (b,c) are at $V_{Z_c}=7.06$ meV; the cyan dashed lines in (b,c) are at $V_z=2.05$ meV, where the ZBCP firstly starts; the first purple dashed lines in (b,c) are at $V_z=2.5$ meV; the second purple dashed lines in (b,c) are at $V_z=8.0$ meV.}
	\label{fig:noSelf11}
\end{figure*}

\begin{figure*}[htb]
	\includegraphics[scale=0.17]{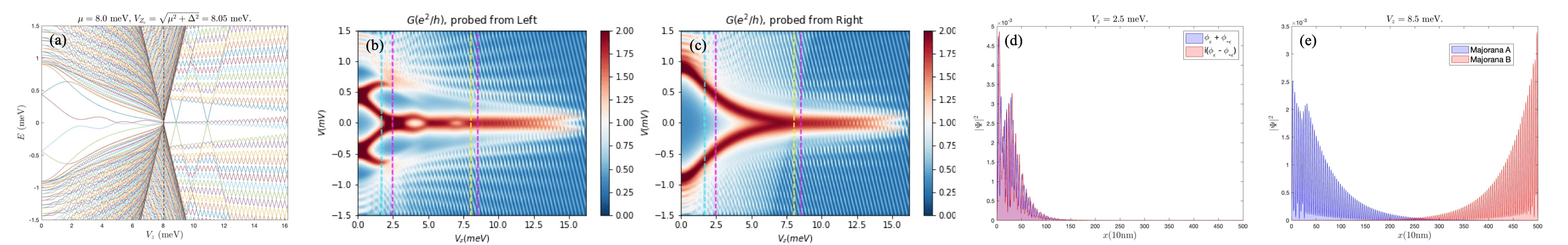}
	\caption{Energy spectrum, conductance measured from both ends, and the lowest lying wave functions below and above $V_{Z_c}=8.05$ meV, for the case of chemical potential $\mu=8.0$ meV. (a): energy spectrum. (b): conductance $G(V)$ measured from the left lead. (c): conductance $G(V)$ measured from the right lead. (d): lowest lying wave functions in the nanowire at $V_z=2.5$ meV, below $V_{Z_c}$. (e): lowest lying wave functions in the nanowire at $V_z=8.5$ meV, above $V_{Z_c}$. The superconducting collapsing field is at $V_c=16.4$ meV. The black dashed line in (a) and yellow dashed lines in (b,c) are at $V_{Z_c}=8.05$ meV; the cyan dashed lines in (b,c) are at $V_z=1.7$ meV, where the ZBCP firstly starts; the first purple dashed lines in (b,c) are at $V_z=2.5$ meV; the second purple dashed lines in (b,c) are at $V_z=8.5$ meV.}
	\label{fig:noSelf12}
\end{figure*}

\begin{figure}[hbt]
	\includegraphics[scale=0.25]{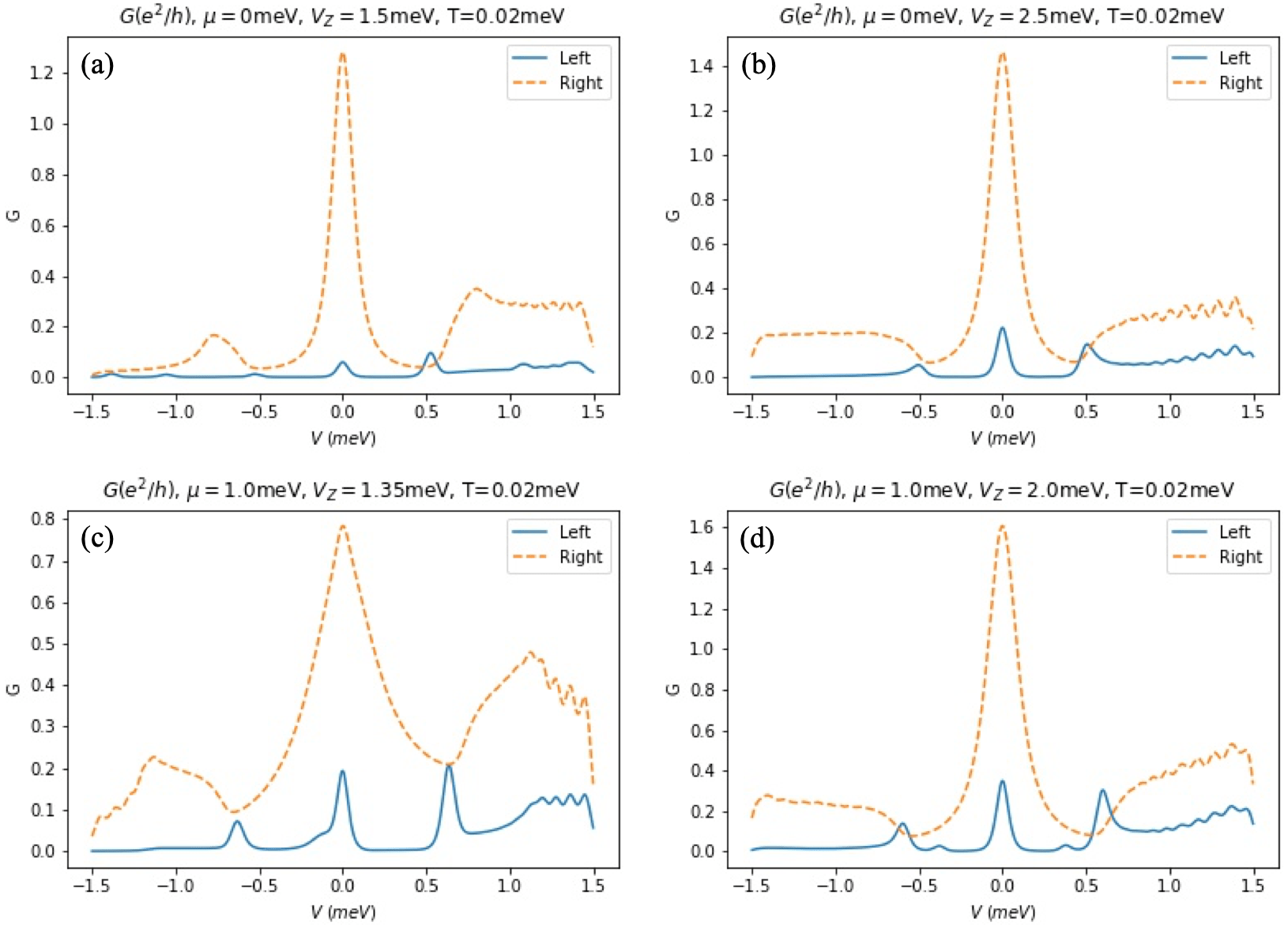}
	\caption{Line cuts of the tunneling conductance as a function of bias voltage from both ends (The blue solid line is from the left end, while the orange dashed line is from the right end.), for low values of chemical potential $\mu$ and $V_z\geq V_{Z_c}$. (a): $\mu=0$ meV and $V_z=1.5$ meV, with $V_{Z_c}=0.90$ meV. (b): $\mu=0$ meV and $V_z=2.5$ meV, with $V_{Z_c}=0.90$ meV. (c): $\mu=1.0$ meV and $V_z=1.35$ meV, with $V_{Z_c}=1.35$ meV. (d): $\mu=1.0$ meV and $V_z=2.0$ meV, with $V_{Z_c}=1.35$ meV. Note that the superconducting collapsing field is fixed at $V_c=20$ meV. }
	\label{fig:LineCut1}
\end{figure}

\begin{figure}[hbt]
	\includegraphics[scale=0.25]{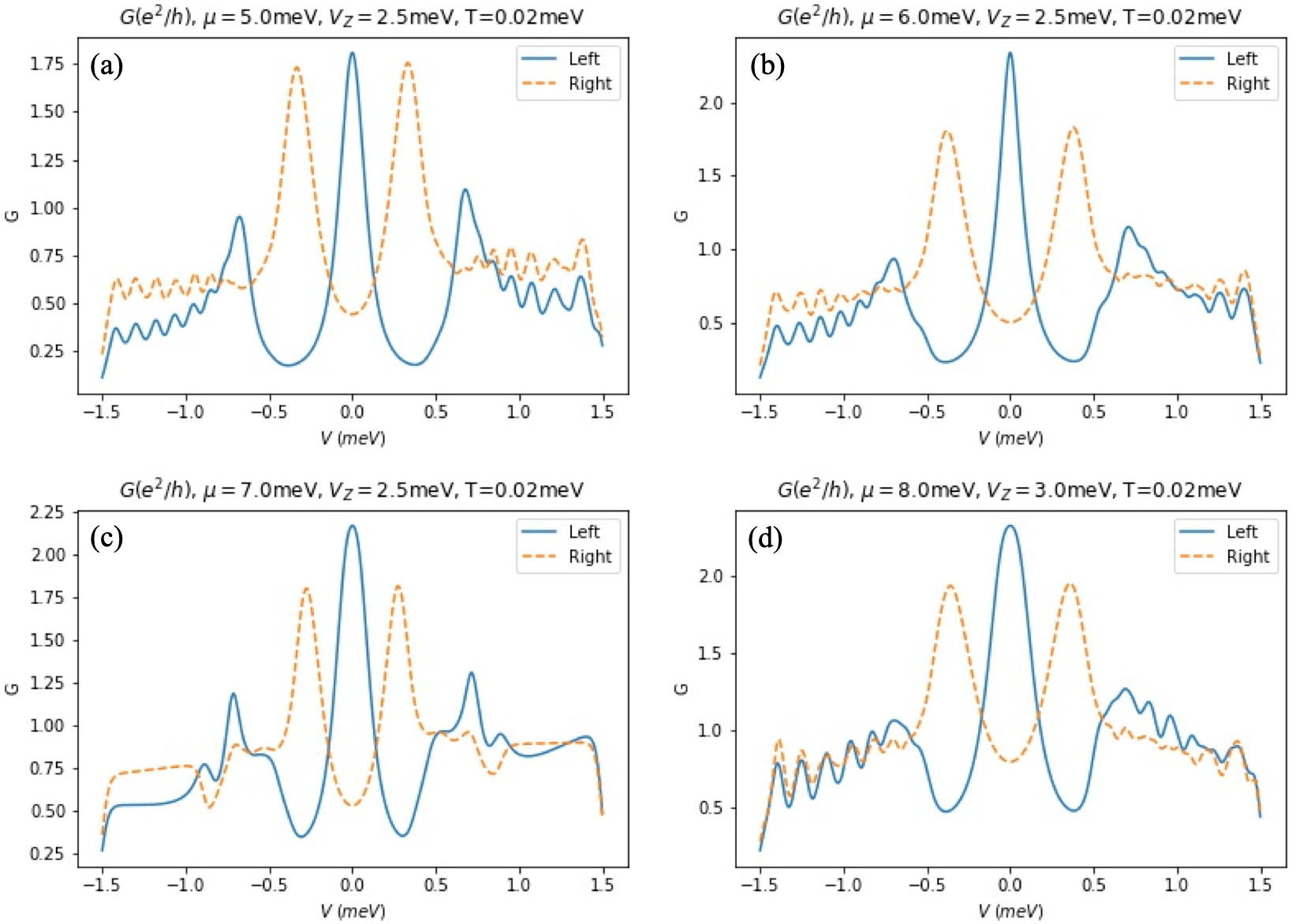}
	\caption{Line cuts of the tunneling conductance as a function of bias voltage from both ends (The blue solid line is from the left end, while the orange dashed line is from the right end), for high values of chemical potential $\mu$ and low values of magnetic field $V_z(<V_{Z_c})$. (a): $\mu=5.0$ meV and $V_z=2.5$ meV, with $V_{Z_c}=5.08$ meV. (b): $\mu=6.0$ meV and $V_z=2.5$ meV, with $V_{Z_c}=6.07$ meV. (c): $\mu=7.0$ meV and $V_z=2.5$ meV, with $V_{Z_c}=7.06$ meV. (d): $\mu=8.0$ meV and $V_z=3.0$ meV, with $V_{Z_c}=8.05$ meV. Note that the superconducting collapsing field is fixed at $V_c=20$ meV.}
	\label{fig:LineCut2}
\end{figure}

\begin{figure*}[hbt]
	\includegraphics[scale=0.175]{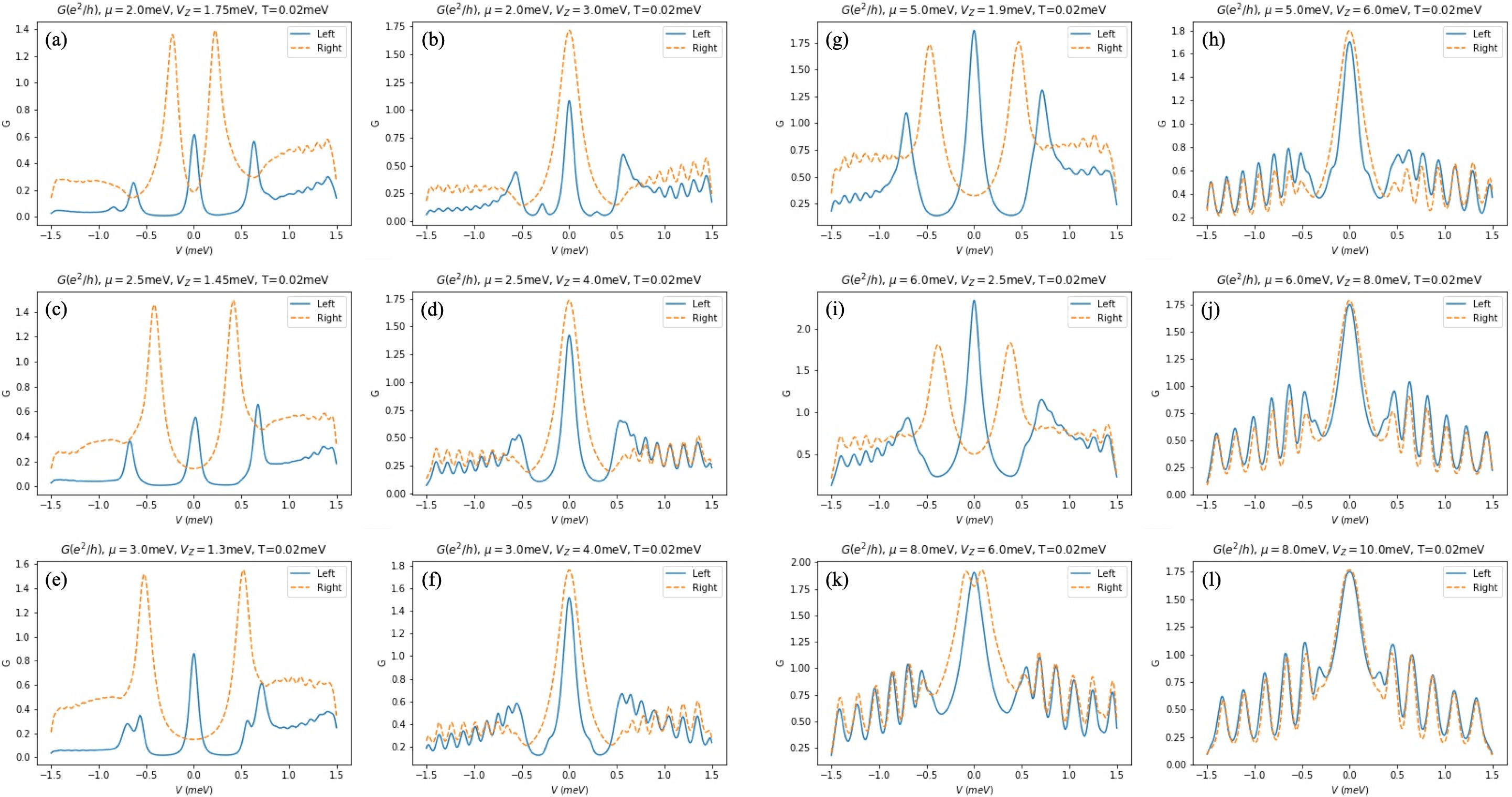}
	\caption{Line cuts of the tunneling conductance as a function of bias voltage from both ends. (The blue solid line is from the left end, while the orange dashed line is from the right end) The magnetic fields $V_z$ for the first and the third columns are set below TQPT point ($V_z<V_{Z_c}$), while $V_z$ for the second and the fourth columns are set above TQPT point ($V_z>V_{Z_c}$). (a): $\mu=2.0$ meV and $V_z=1.75$ meV, with $V_{Z_c}=2.19$ meV. (b): $\mu=2.0$ meV and $V_z=3.0$ meV, with $V_{Z_c}=2.19$ meV. (c): $\mu=2.5$ meV and $V_z=1.45$ meV, with $V_{Z_c}=2.66$ meV. (d): $\mu=2.5$ meV and $V_z=4.0$ meV, with $V_{Z_c}=2.66$ meV. (e): $\mu=3.0$ meV and $V_z=1.3$ meV, with $V_{Z_c}=3.13$ meV. (f): $\mu=3.0$ meV and $V_z=4.0$ meV, with $V_{Z_c}=3.13$ meV. (g): $\mu=5.0$ meV and $V_z=1.9$ meV, with $V_{Z_c}=5.08$ meV. (h): $\mu=5.0$ meV and $V_z=6.0$ meV, with $V_{Z_c}=5.08$ meV. (i): $\mu=6.0$ meV and $V_z=2.5$ meV, with $V_{Z_c}=6.07$ meV. (j): $\mu=6.0$ meV and $V_z=8.0$ meV, with $V_{Z_c}=6.07$ meV. (k): $\mu=8.0$ meV and $V_z=6.0$ meV, with $V_{Z_c}=8.05$ meV. (l): $\mu=8.0$ meV and $V_z=10.0$ meV, with $V_{Z_c}=8.05$ meV. Note that the superconducting collapsing field is fixed at $V_c=20$ meV.}
	\label{fig:LineCut3}
\end{figure*}

\begin{figure*}[hbt]
	\includegraphics[scale=0.2]{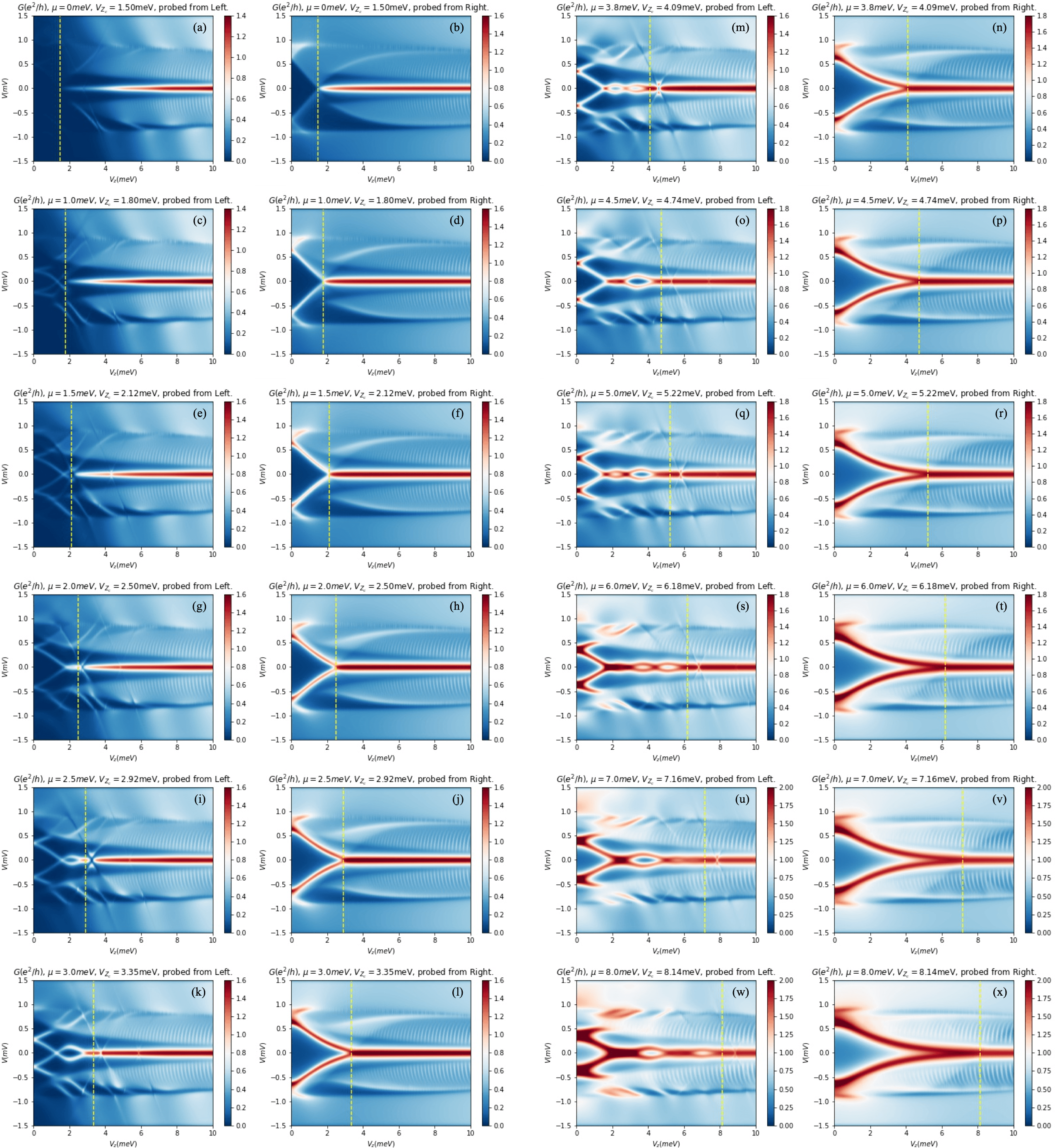}
	\caption{Differential conductance measured from both ends, including the self-energy. The self-energy coupling constant is $\lambda=1.5$ meV, which renormalized the induced superconducting gap at $\omega=0$ to be $\Delta_0=1.5$ meV. Note that the superconducting collapsing field is fixed at $V_c = 20$ meV. The first and third columns are measured from the left lead, while the second and fourth columns are measured from the right lead. The yellow dashed line is where $V_{Z_c}$ marks. (a)-(b): $\mu=0$ meV  and $V_{Z_c}=1.50$ meV. (c)-(d): $\mu=1.0$ meV and $V_{Z_c}=1.80$ meV. (e)-(f): $\mu=1.5$ meV and $V_{Z_c}=2.12$ meV. (g)-(h): $\mu=2.0$ meV and $V_{Z_c}=2.50$ meV. (i)-(j): $\mu=2.5$ meV and $V_{Z_c}=2.92$ meV. (k)-(l): $\mu=3.0$ meV and $V_{Z_c}=3.35$ meV. (m)-(n): $\mu=3.8$ meV and $V_{Z_c}=4.09$ meV. (o)-(p): $\mu=4.5$ meV and $V_{Z_c}=4.74$ meV. (q)-(r): $\mu=5.0$ meV and $V_{Z_c}=5.22$ meV. (s)-(t): $\mu=6.0$ meV and $V_{Z_c}=6.18$ meV. (u)-(v): $\mu=7.0$ meV and $V_{Z_c}=7.16$ meV. (w)-(x): $\mu=8.0$ meV and $V_{Z_c}=8.14$ meV.}
	\label{fig:GSelfE}
\end{figure*}

This section collects the similar numerical results as Fig.\ref{fig:noSelf2} and \ref{fig:noSelf9}, but with different chemical potentials. One of these cases (Fig.~\ref{fig:noSelf1}) for which the chemical potential is smaller such that there is no ZBCP below TQPT. In this case, we only show the lowest-lying wave function below $V_{Z_c}$. For the larger chemical potentials $\mu$, where we can observe ZBCPs both below and above TQPT, we show the lowest-lying wave functions (panel d and e in Figs.\ref{fig:noSelf3}-~\ref{fig:noSelf12}) both below $V_{Z_c}$ (ABS-induced ZBCP) and above $V_{Z_c}$ (MBS-induced ZBCP). In addition, the line cuts of the tunneling conductance are also demonstrated here (Figs.\ref{fig:LineCut1}-\ref{fig:LineCut3}). We also show the differential conductances from both ends of the Majorana nanowire with the self-energy (Fig.~\ref{fig:GSelfE}), illustrating our main idea (the presence of correlation of ZBCPs from both sides of nanowire guarantees the existence of MBS, while one side of ZBCP only results from ABS) doesn't lose even considering the strong-coupling superconductor-semiconductor proximity effect.\\

\clearpage

In Figs.\ref{fig:LineCut1}-\ref{fig:LineCut3}, we show the line cuts of the tunneling conductance from both ends (as a function of bias voltage) at fixed Zeeman splitting values for various chemical potentials, both below (i.e. the trivial regime, $V_z<V_{Z_c}$) and above (i.e. the topological regime, $V_z>V_{Z_c}$) TQPT so that the tunneling physics of both ABS and MBS are manifest, clearly demonstrating the role of end-to-end correlations in determining the identity of Majorana zero modes. In Fig.\ref{fig:LineCut1}, we show right and left tunneling conductance comparisons for low values of chemical potential, $\mu=0$ and 1 meV. Here, the ZBCPs are all induced by MBS, occurring at $V_z>V_{Z_c}$, and as such, produce correlated conductance peaks at the same $V_z$ from both ends, although the peak conductances are very different for tunneling from left and right ends for the chosen parameter values.

\clearpage

In contrast to Fig.~\ref{fig:LineCut1} which presents conductance line cuts for low values of chemical potential, we show in Fig.~\ref{fig:LineCut2} comparative left and right conductance line cuts for large chemical potentials ($\mu=5,6,7,8$ meV all much larger than the induced superconducting gap~1 meV) at low values of $V_z (<V_{Z_c})$ in the trivial regime below TQPT. In each panel of Fig.~\ref{fig:LineCut2}, there is a prominent ABS-induced ZBCP in the left conductance, but not in the right, clearly establishing that the very prominent left ZBCP in these results must necessarily be ABS-induced features since these left ZBP do not correlate with any corresponding right ZBCP at the same Zeeman field. We note that the ABS-induced ZBCPs from the left in Fig.~\ref{fig:LineCut2} (blue solid curves in Fig.~\ref{fig:LineCut2}) look essentially identical to the MBS-induced ZBCPs from the right in Fig.~\ref{fig:LineCut1} (orange dashed curves in Fig.~\ref{fig:LineCut1}), the only difference between the two situations being that in Fig.~\ref{fig:LineCut2}, there is no corresponding ZBCP from the right (i.e. the orange dashed curves in Fig.~\ref{fig:LineCut2} show basically zero conductance), whereas the corresponding left conductance in Fig.~\ref{fig:LineCut1} do manifest ZBPs (i.e. the blue solid curves in Fig.~\ref{fig:LineCut1} have small discernible peaks correlated with the prominent ZBCPs from the right).

~\\

In Fig.~\ref{fig:LineCut3}, we show extensive comparisons between right and left conductance at fixed magnetic field values for different chemical potentials ($\mu=2$ to 8 meV). In each case, we show calculated examples of both trivial ZBCPs caused by ABS existing (below TQPT) only for tunneling from left (and not from right), as well as correlated ZBCPs caused by MBS in the topological regime (above TQPT) existing in tunneling from both ends.  The presence (absence) of end-to-end conductance peak correlations manifesting in MBS (ABS) is obvious in Fig.~\ref{fig:LineCut3}, clearly bringing out the key message of our work. For MBS-induced ZBCP, the conductance magnitudes from right and left tunneling should become increasingly equal as the magnetic field increases, particularly for large values of chemical potential, as can be seen in Fig.~\ref{fig:LineCut3}.

\clearpage

~\\

In Fig.\ref{fig:GSelfE}, we show the calculated comparative conductance plots from two wire ends (similar to Figs.\ref{fig:noSelf2}, \ref{fig:noSelf9}, and \ref{fig:noSelf1}-\ref{fig:noSelf12}) for the strong-coupling superconductor-semiconductor proximity model, where the tunnel coupling between the parent superconductor and the nanowire is large so that the induced superconducting gap in the nanowire is now limited by the parent gap and not by the tunneling at the interface (as it is in the weak coupling results presented so far in Figs.\ref{fig:noSelf2}, \ref{fig:noSelf9}, and \ref{fig:noSelf1}-\ref{fig:LineCut3}, where the proximity gap in the nanowire is smaller than the parent gap). The strong coupling model must necessarily incorporate the dynamical self-energy effects, and the detailed theory has been already provided in the literature \cite{Sau2010Robustness,Stanescu2010Proximity,Sau2012Experimental,Liu2017Andreev}, which we do not reproduce here. The calculated tunneling conductance results in Fig.~\ref{fig:GSelfE} for various chemical potentials show the same qualitative features as what is described above for the weak-coupling results presented in Figs.\ref{fig:noSelf2}, \ref{fig:noSelf9}, and \ref{fig:noSelf1}-\ref{fig:LineCut3}. In particular, the MBS-induced ZBCPs for $V_z>V_{Z_c}$ always manifest together in a correlated manner from both ends of the wire, whereas the ABS-induced ZBCPs for $V_z<V_{Z_c}$ manifest only for tunneling from the left end without manifesting any correlations. Thus, our conclusion about the presence (absence) of end-to-end correlations in the ZBCP implying the presence of topological MBS (trivial ABS) in the system remains valid in the presence of strong-coupling proximity effect.  Note that there are quantitative differences between the two approximations in the details of conductance magnitudes and TQPT points and so on as one would expect, but the key point of ZBCP correlations from the two ends being a decisive signature for MBS is equally applicable to both models.

\end{document}